# A new version of fermion coupled coherent states method: Theory and applications in simulation of two-electron systems

Mohammadreza Eidi[1, 2], Mohsen Vafaee[1*], Ali Reza Niknam[2] and Nader Morshedian[3]

[1] *Department of Chemistry, Tarbiat Modares University, P.O.Box 14115-175, Tehran, Iran*

[2] *Laser and Plasma Research Institute, Shahid Beheshti University, G.C., Tehran, Iran and*

[3] *Research School of Plasma and Nuclear Fusion, NSTRI, P.O.Box 14399-51113, Tehran, Iran*

We report a new version of fermion coupled coherent states method (FCCS-II) to simulate two-electron systems based on a self-symmetrized six-dimensional (6D) coherent states grid. Unlike the older fermion coupled coherent states method (FCCS-I), FCCS-II does not need any new equations in comparison with the coupled coherent states method. FCCS-II uses a simpler and more efficient approach for symmetrizing the spatial wave function in the simulation of fermionic systems. This method, has significantly increased the speed of computations and give us the capability to simulate the quantum systems with the larger CS grids. We apply FCCS-II to simulate the Helium atom and the Hydrogen molecule based on grids with a large numbers of coherent states. FCCS-II with a relatively low number of CS gives a potential energy curve for $H_2$ that is very close to the exact potential curve. Moreover, we have re-derived all the important equations of the FCCS-I method.

## I. INTRODUCTION

During the last two decades, the coupled coherent states (CCS) method has been developed for simulating the quantum dynamics of high-dimensional systems by solving the time dependent Schrödinger equation (TDSE) in the phase space based on coherent states [1-13]. Two of other trajectory-guided approaches are the variational multi-configurational Gaussian approach (vMCG) [7,14-16] and the multiple spawning (MS) method [7,16,17]. vMCG uses time-dependent Gaussian functions as the basis set. Basis sets which obey vMCG equations, do not follow classical trajectories. The equations of vMCG are derived from the variational principle [7]. Therefore, vMCG potentially have the ability to get the best possible solution of the Schrödinger equation. Although, the vMCG equations are complicated and numerically expensive, but vMCG method is able to directly describe quantum events such as tunneling and passage through a conical intersection and at the same time the convergence is fast. vMCG method provides a good description of the dynamics of a molecular system using only a small basis set and subsequently a small number of parameters [16]. MS uses a quantum mechanical wavepacket described by a superposition of Gaussian basis functions that unlike vMCG follows classical trajectories. Hence, MS would not be a good choice for simulating two-electron systems. Also, MS have a great ability to manage the size of the basis set when required [16,18]. As for MS, the vMCG method has been successfully applied in the context of non-adiabatic photochemistry and it appears to be a quite reliable, efficient and cheap approach to deal with non-adiabatic transitions between coupled electronic states while keeping the advantage of calculating the potential energy surfaces (PES) and non-adiabatic couplings on-the-fly [16]. The CCS methodology is situated between vMCG and MS. The CCS method has many considerable advantages which distinguish it from other trajectory guided approaches. The main advantage is that fewer configurations are needed for simulating a system with large number of degrees of freedom. Another advantage is that, the singularity of the Coulombic potentials can be removed and replaced by an error function [5,9]. For more information about the main concepts of the CCS method see Refs. [4,7].

The investigation of non-perturbative laser induced phenomena in many-electron atoms and molecules, such as non-sequential double ionization (NSDI) and high-order harmonic generation (HHG) has formed a growing area of research [19,20]. In multi-electron atoms, He provides the only conceivable meeting ground between ab initio theory and experiment in multiple ionization of atoms by the intense laser fields [19]. Simulation of He exposed to an intense laser field with the wavelength near 800 nm (most frequently used in experiments) and comparison of simulation results with the experiments has not been accomplished yet [19]. Some ab initio TDSE calculations beyond the one dimensional (1D) approximation for the interaction of He and $H_2$ with intense few-cycle near-infrared laser pulses have been reported by Parker et al. [21] and Ruiz et al. [22], respectively. Belfast group performed this comparison under a linearly polarized laser field for a shorter wavelength at 390 nm [21]. However, most computations are carried out for two-electron systems considering one dimension for each electron, and with classical [23-26] and quantum [27] nuclear dynamics. Full dimensional study of two-electron systems like He and $H_2$ in the presence of an intense laser field is not possible yet. By developing the CCS method, it is hoped that solving this major problem become possible.

Originally, the CCS method has been developed to treat the motion of distinguishable particles. For simulating the dynamics of fermion particles, a modified version of the CCS method has been introduced by Kirrander and Shalashilin as fermion coupled coherent states (FCCS) [11]. We have labeled this Shalashilin's approach as FCCS-I throughout this article. The CCS and FCCS-I methods have tried to develop a useful tool for simulating atomic and molecular systems in full dimensions and investigating the dynamics of electrons in systems interacting with intense laser radiation and related phenomena. Some simulations performed by CCS and FCCS-I methods can be listed as follows: the 6D simulation of $H_2$ and its electronic states by the standard CCS [5,6], simulation of He double ionization [9], the strong-field ionization of He at long wavelengths [10], electron dynamics in the laser fields by the FCCS-I method on the base of Frozen Gaussians [11], high-order harmonic generation by the CCS approach [13] and other reported applications [28-34]. Another work done by Z. Zhou and S.-I. Chu [12] is the full dynamics of $H_2$ in intense linearly polarized laser fields which in fact used the Heller's Frozen Gaussians method instead of the CCS method.

---

\* *Corresponding author: m.vafaee@modares.ac.ir*

In the simulations reported on the basis of the CCS and the FCCS-I methods, high energy coherent states are excluded from the grid. Therefore, the grid is biased to the regions with the lowest energy [6]. Furthermore, the diffusion Monte-Carlo (DMC) method on the basis of these two methods has needed a grid refinement algorithm like the maximizing the residual overlap (MRO) [6]. Moreover, the FCCS-I method uses a symmetrizing equation that makes equations complex and computations cumbersome. Here, we introduce a new version of fermion coupled coherent states method. We have labeled this version of FCCS as FCCS-II throughout this article. This new version of FCCS method does not need to use any additional symmetrizing equation, biasing the grid to the regions with the lowest energy and any grid refinements.

In this article, after giving a brief review on the CCS and the FCCS-I methods, we introduce FCCS-II method. In Sec II A, coherent states and the CCS method have been investigated and reviewed in a new manner. In Sec II B, we have studied the FCCS-I method and proposed FCCS-II method. Moreover, in Sec II B, we have employed a new random coherent states grid generation method which considers two compression parameters for position and momentum coordinates in the phase space. In Sec II C, the diffusion Monte-Carlo and imaginary time propagation methods has been introduced for FCCS-II method. Finally, we have applied the FCCS-II method to simulate the ground state of He and the potential well of $H_2$ in Sec III.

## II. THEORY
### A. The coupled coherent states method (CCS)

We give a brief review of basics of coherent states and the CCS method. In this part, some important equations of the CCS method [4,7]. will be re-derived. Coherent states (CS) are eigenkets of the annihilation operator $\hat{a}$ and eigenbras of the creation operator $\hat{a}^\dagger$ as

$$a|Z\rangle \equiv Z|Z\rangle \qquad (1)$$
$$\langle Z|a^\dagger \equiv \langle Z|Z^*$$

where eigenvalue $Z$ has this form

$$Z = \frac{\gamma^{1/2}}{\sqrt{2}}q + i\frac{\gamma^{-1/2}}{\sqrt{2}\hbar}p \qquad (2)$$

where $q$ is the position and $p$ is the momentum of the wave packet with fixed coordinate space width $\gamma$. We can name this CS as standard (asymmetric) coherent state (ACS) compared to the symmetrized coherent states that will be named as SCS in the next section. Coherent states construct a nonorthogonal overcomplete basis set as

$$\langle Z|Z'\rangle = \prod_{j=1}^{M} exp\left(-\frac{1}{2}\left(|z_j|^2 + |z_j'|^2\right) + z_j^* z_j'\right). \qquad (3)$$

For a two-electron system, $M$, the number of dimensions, is equal to six. The Hamiltonian operator $\hat{H}(\hat{P}, \hat{Q})$ can be expressed in the terms of the creation and the annihilation operators $\hat{H}(\hat{a}, \hat{a}^\dagger)$. The Hamiltonian operator $\hat{H}(\hat{a}, \hat{a}^\dagger)$, can be reordered in such a way that all creation operators place on the left $\tilde{H}(\hat{a}^\dagger, \hat{a})$. The matrix elements of the ordered Hamiltonian operator $\tilde{H}(\hat{a}^\dagger, \hat{a})$ can be easily derived by the use of Eq. (1). Then, we have

$$\langle Z|\tilde{H}(\hat{a}^\dagger, \hat{a})|Z'\rangle = \langle Z|Z'\rangle\tilde{H}(Z^*, Z'). \qquad (4)$$

Identity operator of coherent states has this form

$$\mathbb{I} = \sum_{k,l=1}^{N} |Z_k\rangle(\Omega^{-1})_{kl}\langle Z_l|. \qquad (5)$$

where $N$ is the number of CS. In Eq. (5), $\Omega^{-1}$ is the inverse of the overlap matrix $\Omega$ with elements

$$\Omega_{kl}(t) = \langle Z_k(t)|Z_l(t)\rangle. \qquad (6)$$

In the coordinate representation, these $M$ dimensional coherent states are Gaussian wave packets with fixed width $\gamma$ [8]

$$\langle X|Z\rangle = \prod_{j=1}^{M}\left(\frac{\gamma_j}{\pi}\right)^{1/4} exp\left(-\frac{\gamma_j}{2}\left(x^{(j)} - q^{(j)}\right)^2 \right. \qquad (7)$$
$$\left. + \frac{i}{\hbar}p^{(j)}\left(x^{(j)} - q^{(j)}\right) + \frac{ip^{(j)}q^{(j)}}{2\hbar}\right).$$

Wave function of a system with $M$ degrees of freedom can be represented as a superposition of $N$ trajectory-guided coherent states

$$|\psi(t)\rangle = \sum_{k=1}^{N} D_k(t) exp\left(i\frac{S_k(t)}{\hbar}\right)|Z_k(t)\rangle. \qquad (8)$$

This is the main idea of the CCS method [4,7,11]. In the Eq. (8), preexponential factor $D_k(t)$ can be derived by

$$D_k(t) = \sum_{l=1}^{N} C_l(t) exp\left(i\frac{(S_l(t) - S_k(t))}{\hbar}\right)(\Omega^{-1})_{kl}(t) \qquad (9)$$

where

$$C_l(t) = \langle Z_l(t)|\psi(t)\rangle exp\left(-\frac{i}{\hbar}S_l(t)\right). \qquad (10)$$

In Eqs. (9) and (10), $S(t)$ is the classical action

$$S(t) = \int \ell \, dt. \quad (11)$$

where $\ell$ is the diagonal matrix elements of the Lagrangian operator in the representation of coherent states

$$\hat{\mathcal{L}} = \frac{i\hbar}{2}\left(\frac{\vec{\partial}}{\partial t} - \frac{\overleftarrow{\partial}}{\partial t}\right) - \widehat{H}. \quad (12)$$

On the base of ACS, the diagonal matrix elements of the Lagrangian operator can be obtained as follows

$$\ell = \langle Z|\hat{\mathcal{L}}|Z\rangle = \frac{i\hbar}{2}(\langle Z|\dot{Z}\rangle - \langle \dot{Z}|Z\rangle) - \langle Z|\widehat{H}|Z\rangle. \quad (13)$$

For the dynamic equation of the classical action $dS(t)/dt$ on the base of ACS, it is needed to compute $\langle Z|\dot{Z}\rangle$ and $\langle \dot{Z}|Z\rangle$. The CS $|Z'\rangle$ can be generated from the vacuum state $|0\rangle$ as follows

$$|Z'\rangle = exp\left(Z'\hat{a}^\dagger - \frac{1}{2}Z'Z'^*\right)|0\rangle \quad (14)$$
$$= \prod_{j=1}^{M} exp\left(Z'_j \hat{a}_j^\dagger - \frac{1}{2}Z'_j Z'^*_j\right)|0\rangle.$$

Time derivative of $|Z'\rangle$ is derived as follows

$$|\dot{Z}'\rangle = \sum_{j=1}^{M}\left(\frac{\partial |Z'\rangle}{\partial Z'_j}\frac{\partial Z'_j}{\partial t} + \frac{\partial |Z'\rangle}{\partial Z'^*_j}\frac{\partial Z'^*_j}{\partial t}\right). \quad (15)$$

From Eq. (14), one can see that

$$\frac{\partial |Z'\rangle}{\partial Z'_j} = \left(\hat{a}_j^\dagger - \frac{1}{2}Z'^*_j\right)|Z'\rangle + \frac{\partial \hat{a}_j^\dagger}{\partial Z'_j}Z'|Z'\rangle \quad (16)$$

$$\frac{\partial |Z'\rangle}{\partial Z'^*_j} = \left(-\frac{1}{2}Z'_j\right)|Z'\rangle + \frac{\partial \hat{a}_j^\dagger}{\partial Z'^*_j}Z'|Z'\rangle. \quad (17)$$

By substitution of these two equations in Eq. (15) and knowing that

$$\frac{\partial \hat{a}_j^\dagger}{\partial Z'_j}\frac{\partial Z'_j}{\partial t} + \frac{\partial \hat{a}_j^\dagger}{\partial Z'^*_j}\frac{\partial Z'^*_j}{\partial t} = \frac{\partial \hat{a}_j^\dagger}{\partial t} = 0 \quad (18)$$

for the first term in Eq. (13), one can show that

$$\langle Z|\dot{Z}'\rangle = \sum_{j=1}^{M}\left(\langle Z|\hat{a}_j^\dagger|Z'\rangle \dot{Z}'_j - \frac{1}{2}(Z'^*_j \dot{Z}'_j + Z'_j \dot{Z}'^*_j)\langle Z|Z'\rangle\right). \quad (19)$$

Using Eq. (1) then we have

$$\langle Z|\dot{Z}'\rangle = \langle Z|Z'\rangle \sum_{j=1}^{M}\left(Z_j^* \dot{Z}'_j - \frac{1}{2}(Z'^*_j \dot{Z}'_j + Z'_j \dot{Z}'^*_j)\right). \quad (20)$$

Consider $Z' = Z$, we can write

$$\langle Z|\dot{Z}\rangle = \frac{1}{2}\sum_{j=1}^{M}(Z_j^* \dot{Z}_j - Z_j \dot{Z}_j^*). \quad (21)$$

Similarly, for the second term in Eq. (13), we will have

$$\langle \dot{Z}|Z'\rangle = \langle Z|Z'\rangle \sum_{j=1}^{M}\left(Z'_j \dot{Z}_j^* - \frac{1}{2}(Z_j \dot{Z}_j^* + Z_j^* \dot{Z}_j)\right) \quad (22)$$

Again by considering $Z' = Z$, we can derive

$$\langle \dot{Z}|Z\rangle = \frac{1}{2}\sum_{j=1}^{M}(Z_j \dot{Z}_j^* - Z_j^* \dot{Z}_j). \quad (23)$$

At the end by substituting Eqs. (21) and (23) in Eq. (13), the dynamic equation of the classical action can be obtained as follows

$$\frac{dS(t)}{dt} = \ell = \frac{i\hbar}{2}\sum_{j=1}^{M}(Z_j^* \dot{Z}_j - Z_j \dot{Z}_j^*) - \widetilde{H}(Z^*, Z). \quad (24)$$

Based on the CCS method, classical trajectories are determined by the classical dynamic equation with quantum corrections [4,7]

$$\frac{\partial Z}{\partial t} = -\frac{i}{\hbar}\frac{\partial \widetilde{H}(Z^*, Z)}{\partial Z^*} \quad (25)$$

and its complex conjugate. This is one of the most important features of the CCS method which distinguishes it from similar methods [7]. In Eq. (25), $\widetilde{H}(Z^*, Z)$ is a diagonal element of the Hamiltonian matrix. Applying coherent states identity operator, the TDSE can be represented in the CS basis set as

$$\langle Z_j|\frac{d|\psi\rangle}{dt} = -\frac{i}{\hbar}\sum_{k,l=1}^{N}\langle Z_j|\widehat{H}|Z_k\rangle(\Omega^{-1})_{kl}\langle Z_l|\psi\rangle. \quad (26)$$

Using Eqs. (4)-(6), (8)-(10), (24) and (25) the evaluation equation for the wave-function can be obtained as

$$\frac{dC_j}{dt} = -\frac{i}{\hbar}\sum_{k}^{N}\langle Z_j|Z_k\rangle \, \delta^2 \widetilde{H}(Z_j^*, Z_k) D_k \, exp\left(\frac{i(S_k - S_j)}{\hbar}\right) \quad (27)$$

where

$$\delta^2 \widetilde{H}(Z_j^*, Z_k) = \widetilde{H}(Z_j^*, Z_k) - \widetilde{H}(Z_j^*, Z_j) \\ - \frac{\partial \widetilde{H}(Z_j^*, Z_j)}{\partial Z_j}(Z_k - Z_j). \quad (28)$$

The initial points for solving Eq. (27) are

$$C_l(0) = \langle z_l(0)|\psi(0)\rangle. \quad (29)$$

Time dependent expectation value of every observable can be computed by using following formula

$$\langle \psi|\hat{O}|\psi\rangle = \sum_{i,j=1}^{N}\langle Z_i|\hat{O}|Z_j\rangle D_i^* D_j exp\left(i\frac{S_j - S_i}{\hbar}\right), \quad (30)$$

where $\langle Z_i|\hat{O}|Z_j\rangle$ is the matrix elements of a typical $\hat{O}$ observable. For example, Eq. (30) could be used to compute the expectation value of the Hamiltonian of two-electron systems such as $H_2$ [4,5]. Consider fixed nucleus in space, the Hamiltonian of a Hydrogen like molecules will be

$$H = \frac{p_{e1}^2}{2} + \frac{p_{e2}^2}{2} - \frac{1}{|r_{e1} - R_1|} - \frac{1}{|r_{e1} - R_2|} - \frac{1}{|r_{e2} - R_1|} \quad (31)$$
$$- \frac{1}{|r_{e2} - R_2|} + \frac{1}{|r_{e2} - r_{e1}|}.$$

Hence, we have to calculate the matrix elements of each terms of this equation. The matrix elements of the electron-nuclei Coulombic potentials on the base of ACS, can be derived as [5,9]

$$\left\langle Z \left| \frac{1}{|r_{ei} - R_k|} \right| Z' \right\rangle = \langle Z|Z' \rangle \frac{1}{\sqrt{\rho_{i(k)}^2}} erf\left(\sqrt{\gamma \rho_{i(k)}^2}\right) \quad (32)$$

where

$$\boldsymbol{\rho}_{i(k)} = \frac{Z_{ei}^* + Z_{ei}'}{\sqrt{2\gamma}} - \boldsymbol{R}_k \quad (33)$$

and the matrix elements of the electron-electron Coulombic potential can be derived as [5,9]

$$\left\langle Z \left| \frac{1}{|r_{e1} - r_{e2}|} \right| Z' \right\rangle = \langle Z|Z' \rangle \frac{1}{\sqrt{\rho_{12}^2}} erf\left(\sqrt{\frac{\gamma}{2} \rho_{12}^2}\right) \quad (34)$$

where

$$\rho_{12} = \frac{Z_{e1}^* + Z_{e1}'}{\sqrt{2\gamma}} - \frac{Z_{e2}^* + Z_{e2}'}{\sqrt{2\gamma}}. \quad (35)$$

Both Eq. (32) and Eq. (34) show that Coulombic singularities have been replaced by the complex error function $erf$. This is also one of the main advantages of the CCS method. For a two-electron He like atom, nuclei-nuclei Coulombic potential term is eliminated and $R_1 = R_2 = 0$ in Eq. (31).

**B. Fermion Coupled Coherent State method**

Based on the Fermi-Dirac statistics, wave function of a fermionic system is antisymmetric under interchange of any pair of particle labels. Two-electron wave function is combined from spin and spatial wave functions. Based on the Pauli exclusion principle, no two identical fermions can be in the same quantum state. For example, in the ground state, both electrons have symmetric spatial wave function. Hence, they must have an antisymmetric spin wave function (singlet spin state). In an excited state, both electrons can have antisymmetric spatial wave function corresponding to the triplet state. To simulate fermionic systems by the CCS method, following symmetric or antisymmetric 6D coherent states has been proposed by Shalashilin and Child [5] and Kirrander and Shalashilin [11] . (As mentioned before, we have labeled this version of fermion coupled coherent states method as FCCS-I)

$$|Z_{S/AS}\rangle = \frac{|Z_{e1}Z_{e2}\rangle \pm |Z_{e2}Z_{e1}\rangle}{\sqrt{2(1 \pm a)}} \quad (36)$$

where

$$a = |\langle Z_{e1}|Z_{e2}\rangle|^2. \quad (37)$$

When spatial wave function is symmetric (singlet state), it is necessary to take symmetrized coherent state $|Z_s\rangle$ (SCS). Otherwise, the antisymmetrized coherent state $|Z_{As}\rangle$ must be used for simulating the system. On the base of SCS, it can be verified that the dynamic equation of the classical action becomes

$$\frac{\partial S_s}{\partial t} = \ell_s = \langle Z_s|\hat{L}|Z_s\rangle \quad (38)$$

$$= \frac{i\hbar}{2(1+a)} \sum_{j=1}^{M/2} \left( (Z_j^* + aZ_{j+M/2}^*)\dot{Z}_j \right.$$
$$- (Z_j + aZ_{j+M/2})\dot{Z}_j^*$$
$$+ (Z_{j+M/2}^* + aZ_j^*)\dot{Z}_{j+M/2}$$
$$\left. - (Z_{j+M/2} + aZ_j^*)\dot{Z}_{j+M/2}^* \right)$$
$$- \langle Z_s|\hat{H}|Z_s\rangle.$$

This equation and its elements are re-derived by an exact manner in Appendices A and B. In the Appendix C, we have presented the dynamic equation of the wave function on the basis of FCCS-I. From Eqs. (C18) and (C19), one can verify that the dynamic equation of $C_{sj}$ coefficients have the following form

$$\frac{dC_{sj}}{dt} = -\frac{i}{\hbar} \sum_{k=1}^{N} \delta_s^2 H \, D_{sk} \, exp\left(\frac{i(S_{sk} - S_{sj})}{\hbar}\right). \quad (39)$$

Where

$$\delta_s^2 H = \langle Z_{sj}|\hat{H}|Z_{sk}\rangle - \langle Z_{sj}|\hat{H}|Z_{sj}\rangle\langle Z_{sj}|Z_{sk}\rangle \tag{40}$$

$$+ \sum_{i=1}^{M/2} \left(\left(\frac{1}{1+a_j}\langle Z_{sj}|Z_{sk}\rangle Z_{ji} + \frac{a_j}{1+a_j}\langle Z_{sj}|Z_{sk}\rangle Z_{ji+M/2}\right.\right.$$

$$\left.- 2\left(\frac{\langle Z_{12j}|Z_{12k}\rangle Z_{ki} + \langle Z_{21j}|Z_{12k}\rangle Z_{ki+M/2}}{\sqrt{2(1+a_k)2(1+a_j)}}\right)\right)\frac{\partial\langle Z_{sj}|\hat{H}|Z_{sj}\rangle}{\partial Z_{ji}}$$

$$+ \left(\frac{1}{1+a_j}\langle Z_{sj}|Z_{sk}\rangle Z_{ji+M/2} + \frac{a_j}{1+a_j}\langle Z_{sj}|Z_{sk}\rangle\left(\frac{Z_{ji}^* + Z_{ji}}{2}\right)\right.$$

$$\left.\left.- 2\left(\frac{\langle Z_{12j}|Z_{12k}\rangle Z_{ki+M/2} + \langle Z_{21j}|Z_{12k}\rangle Z_{ki}}{\sqrt{2(1+a_k)2(1+a_j)}}\right)\right)\frac{\partial\langle Z_{sj}|\hat{H}|Z_{sj}\rangle}{\partial Z_{ji+M/2}}\right).$$

The time dependent expectation value of any observable on the base of SCS, can be computed by using following formula

$$\langle\psi|\hat{O}|\psi\rangle = \sum_{i,j=1}^{N}\langle Z_{si}|\hat{O}|Z_{sj}\rangle D_{si}^* D_{sj} exp\left(i\frac{S_{sj} - S_{si}}{\hbar}\right). \tag{41}$$

Using Eq. (36), it can be shown that in Eq. (41) $\langle Z_{si}|\hat{O}|Z_{sj}\rangle$ contains four terms

$$\langle Z_{si}|\hat{O}|Z_{sj}\rangle = \frac{1}{\sqrt{2(1+a_i)}}\frac{1}{\sqrt{2(1+a_j)}}[\langle Z_{12i}|\hat{O}|Z_{12j}\rangle \tag{42}$$
$$+ \langle Z_{12i}|\hat{O}|Z_{21j}\rangle + \langle Z_{21i}|\hat{O}|Z_{12j}\rangle$$
$$+ \langle Z_{21i}|\hat{O}|Z_{21j}\rangle].$$

Here, we introduce a new simpler and more efficient approach for symmetrizing spatial wave function in the simulation of the ground state of fermionic systems on the basis of CCS method which is led to a newer version of the fermion coupled coherent states method. As mentioned before, we name this new version of FCCS as FCCS-II. FCCS-II is based on a self-symmetrized 6D CS grid generated in such a way that:

A. The 6D grid points constructed symmetrically with respect to the origin. i.e. for each 6D coherent state $|Z_k\rangle$ in a randomly generated grid, there is another 6D coherent state $|Z_k'\rangle$ created by the inversion of the position and the momentum of the original one

$$|Z_k'\rangle = |-Z_k\rangle. \tag{43}$$

B. The necessary indistinguishability property for two electrons is applied by:
1) From any 6D coherent state $|Z_k\rangle = |Z_{e1_k}\rangle \otimes |Z_{e2_k}\rangle$ in the grid, there is another 6D coherent state $|Z_k'\rangle$ created by interchanging 3D coherent states $|Z_{e1_k}\rangle$ and $|Z_{e2_k}\rangle$ corresponded to the first and the second electron, respectively

$$|Z_k'\rangle = |Z_{e2_k}\rangle \otimes |Z_{e1_k}\rangle. \tag{44}$$

2) Weight of $|Z_k\rangle$ and $|Z_k'\rangle$ in the part B.1 is set to be equal. To apply this condition, we consider

$$C_k' = \langle Z_k'|\psi\rangle = \langle Z_k|\psi\rangle = C_k. \tag{45}$$

Similar structure was applied previously in the exact numerical solution of TDSE in 1D for each electron in the position representation for $H_2$ [23-26]. FCCS-II instead of using SCS (Eq. (36)) and involving complex equations in the FCCS-I (Eqs. (38)-(42)), uses the equations of standard approach (all equations in Sec II A) which are very simpler and easier to solve.

In implementation of FCCS-II, two initial 3D CS ($z_{0_{1j}}$ and $z_{0_{2j}}$) is generated on the locations of each of two nuclei in the phase space. For each coherent state $k$ corresponded to each electron $i$ in each dimension $j$, two random numbers between zero and one ($q_{ijk}$ and $p_{ijk}$) are generated. Therefore, $i, j$ and $k$ show the electron number, the dimension number, and the coherent state number, respectively. Each coherent state is created with the help of the Gaussian distribution function around $z_{0_{ij}}$ by the following formula

$$z_{ijk} = z_{0_{ij}} + \left(\frac{q_{ijk}}{\alpha_q} + i\frac{p_{ijk}}{\alpha_p}\right) \tag{46}$$

where $i = 1,2$ ; $j = 1,2,3$ ; $k = 1,2..., N$. In the Eq. (46), $\alpha_q$ and $\alpha_p$ are compression parameters for the position and the momentum, respectively. These compression parameters modify the width of the basis set distribution. In the previous works on the basis of CCS and the FCCS-I methods [8] it was taken $\alpha_q = \alpha_p$. We have considered two different compression parameters $\alpha_q$ and $\alpha_p$ in this work. With this option, now we can do more modifications on the CS grid.

## C. Diffusion Monte-Carlo and imaginary time propagation methods

Diffusion Monte Carlo (DMC) is a quantum Monte Carlo method applied to solve TDSE of many quantum systems in imaginary time. DMC is an accurate approach for finding the ground state energy of a quantum system. Recently, a version of DMC based on the CCS method have been reported and applied for the simulation of electronic states of $H_2$ [6]. Based on [6], an iterative refinement technique named the maximizing the residual overlap (MRO) is required to improve the quality of the Monte Carlo CS grid. In addition, in the generation and refinement procedures of CS grid, it is necessary to bias the CS grid to the regions with the lowest energy.

Instead of using the DMC method, we have applied the method of imaginary time propagation of the Schrödinger equation (ITPSE) on the basis of FCCS-II to gain the ground states of two-electron systems. One of the main advantages of the ITPSE on the basis of FCCS-II is that we do not need any refinement algorithm like MRO to improve the quality of initial generated CS grid and converge to the exact ground states. Another advantage of FCCS-II is that exclusion of high energy CS from grid and thereby biasing the CS grid to the regions of the lowest energy, is not necessary.

A brief theory of the ITPSE method has been provided as follows. Coherent states are considered to be frozen and not evaluated by time. Therefore, Eq. (26) is reduced to

$$\frac{dC_j}{dt} = -\frac{i}{\hbar} \sum_{k=1}^{N} D_k \langle Z_j | Z_k \rangle \widetilde{H}(Z_j^*, Z_k). \quad (47)$$

This equation were actually suggested before the CCS method by Huber and Heller [35]. Eq. (47) is propagating in imaginary time $d\tau = idt$ until the average value of the energy

$$\frac{\langle \psi | \widehat{H} | \psi \rangle}{\langle \psi | \psi \rangle} = \frac{\sum_{i,j=1}^{N} \langle Z_i | \widehat{H} | Z_j \rangle D_i^* D_j}{\sum_{j=1}^{N} C_j^* D_j} \quad (48)$$

converges to the lowest accessible energy for a given grid i.e. the energy of the ground state of the system.

## III. CALCULATIONS AND RESULTS

In this work, we apply the ITPSE method on the basis of FCCS-II method for calculating the ground state of two-electron systems like $H_2$ and He. Some computations have been previously done for simulating electronic states of $H_2$ by the CCS method based on grids with a limited numbers of CS [6] but here, it is for the first time that the ground state energy of He is computed and reported by a CCS method. In the simulation of the ground state of He, at first, we have investigated the best $\gamma$ parameter for grids with various numbers of CS. Therefore, the ground state energy of He has been computed based on various CS grids with different $\gamma$ parameters. We picked and plotted the results of three of the best $\gamma$ parameters ($\gamma = 1.0$, 1.5 and 1.8) in the FIG. 1. For all calculations in simulation of He, we have found out that $\alpha_q = 1.5$ and $\alpha_p = 10.0$ are the best compression parameters. From the results shown in FIG. 1, it is deduced that for computations based on small grids (< 200 CS), grids with 200 -1000 CS, and the larger grids (> 1000 CS), the best results are corresponding with $\gamma \simeq 1$, 1.5, and 1.8, respectively.

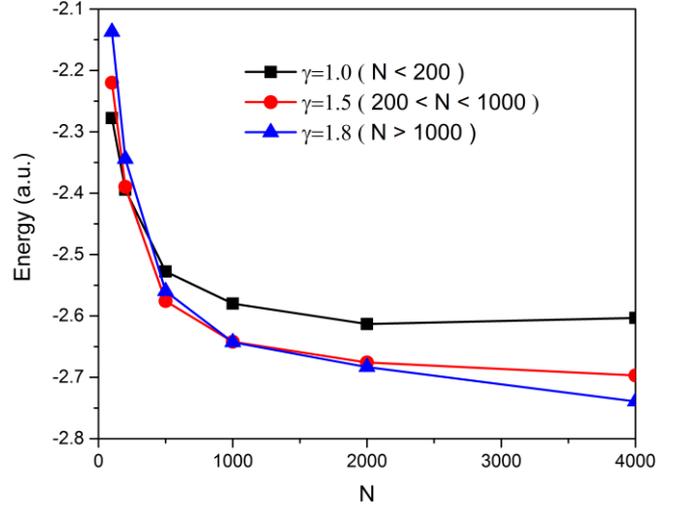

**FIG. 1.** (Color online) The ground state energy of He is computed based on different CS grids for three of the best γ parameters. Each of these γ parameters lead to the best results for the corresponding values in the parenthesis.

At the next step, we have examined the dependency of the energy converged values on $\gamma$ parameter for a grid with 4000 CS. Results for this grid is represented in FIG. 2. This figure shows that $\gamma \simeq 1.8$ would be led to the lowest accessible energy for a grid with 4000 CS.

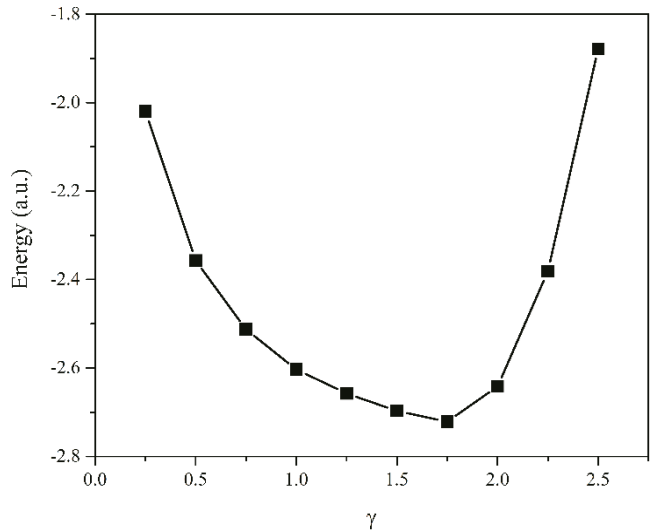

**FIG. 2.** The ground state energy of He computed for different $\gamma$ parameters with 4000 CS.

**Table 1**. The ground state energy of He for different CS grids in comparison with -2.903 the exact value.

| Number of CS | Energy (a.u.) |
|---|---|
| 100 | -2.137 |
| 200 | -2.344 |
| 500 | -2.559 |
| 1000 | -2.642 |
| 2000 | -2.683 |
| 4000 | -2.739 |
| 8000 | -2.765 |
| 10000 | -2.770 |

In the last study on He, the dependency of the computed values for the ground state energy on the number of CS grid points $N$, is investigated. Therefore, the ground state energy of He has been computed based on various grids and the results are shown in **Error! Not a valid bookmark self-reference.** and compared with the exact value.

Now, we apply the ITPSE method on the basis of FCCS-II for computing the well potential of the ground state of $H_2$. In the simulation of the potential well of $H_2$, at first, we looked for the best sets of $\gamma$, $\alpha_q$ and $\alpha_p$ parameters to reach the lowest values. For a grid with 500 CS and a grid with 1000 CS, results are represented in FIG. 3 and FIG. 4, respectively.

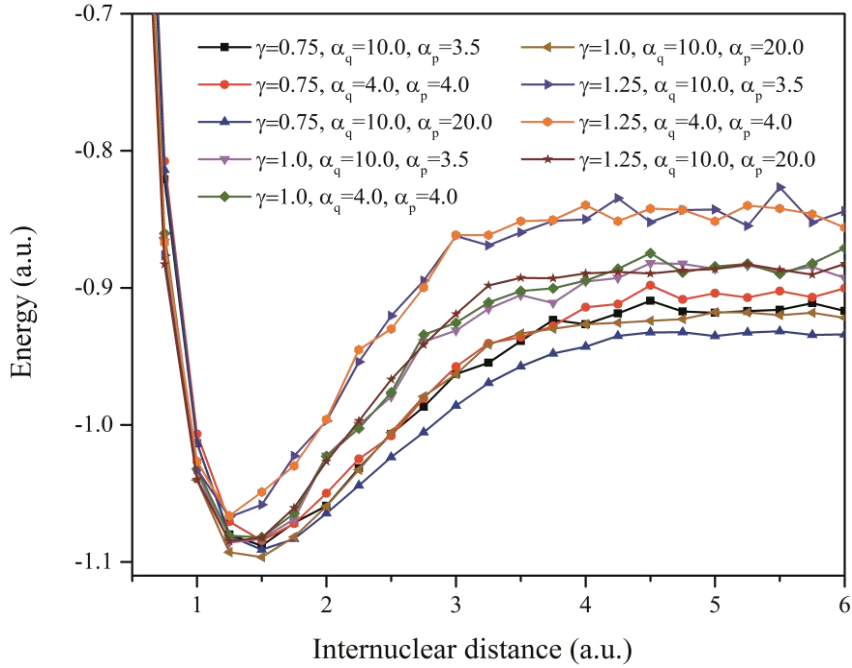

**FIG. 3.** (Color online) The potential well of $H_2$ computed based on a grid with 500 CS for different sets of $\gamma$, $\alpha_q$, and $\alpha_p$ parameters.

FIG. 3 shows that, for $N = 500$ the lowest accessible energy is obtained by $\gamma = 1.0$, $\alpha_q = 10.0$, $\alpha_p = 20.0$ and $\gamma = 0.75$, $\alpha_q = 10.0$, $\alpha_p = 20.0$ for small and large internuclear distances, respectively.

Similarly, FIG. 4 shows that, for $N = 1000$ the lowest accessible energy is achieved by $\gamma = 1.2$ and $\gamma = 0.75$ for small and large internuclear distances, respectively. Also, for grids with 500 and 1000 CS, computations with $\gamma = 0.65$ and $\gamma = 0.85$ are repeated and compared to those of $\gamma = 0.75$. The results show that for large internuclear distances, $\gamma = 0.75$ leads to lower energies. Due to the intrinsic random property of the Monte-Carlo based methods, FCCS-II for grids with a small numbers of CS shows fluctuations in the well potential. These fluctuations decrease by increasing the numbers of CS in the grid. It can be seen in FIG. 3 and FIG. 4 that the fluctuations of 1000 CS is much less than 500 CS. By the way, fluctuations can be decreased by averaging over the results of same repeated computations.

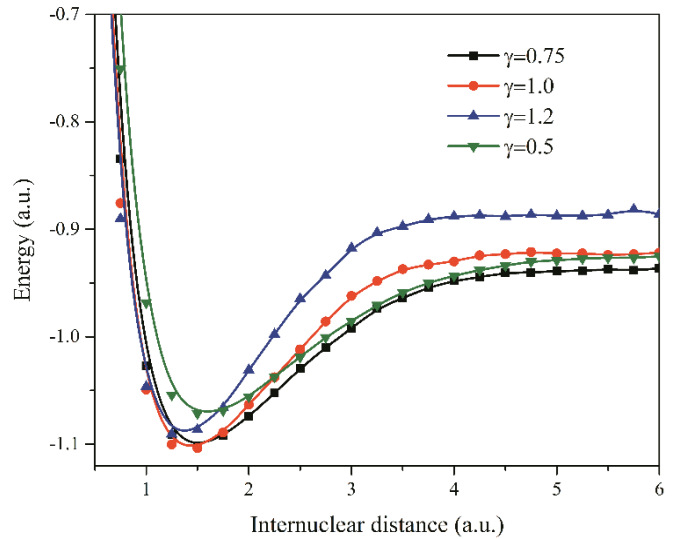

**FIG. 4.** The potential well of $H_2$ computed based on a grid with 1000 CS for different $\gamma$ parameter. Compression parameters $\alpha_q = 10.0$ and $\alpha_p = 3.5$ led to the lowest ground state energy.

Our next aim was to find the dependency of the results on the numbers of CS (*N*) incorporating into the simulation. The CCS method has been applied to compute the ground state of $H_2$ for different numbers of CS up to maximum 400 CS [6] but we have applied FCCS-II for computing the potential well of H2 based on a grid with more than 4000 CS. In FIG. 5, the potential well of H2 has been computed for different grids and compared with the exact values [36-38]. As we expect, this figure shows that for grids involving more coherent states, the computed values would be closer to the exact values. It is very interesting that all curves of FCCS-II on the FIG. 5 are very similar to the exact curve and simply shifted from the exact result. This figure shows that extrapolation allows to get close to the exact result.

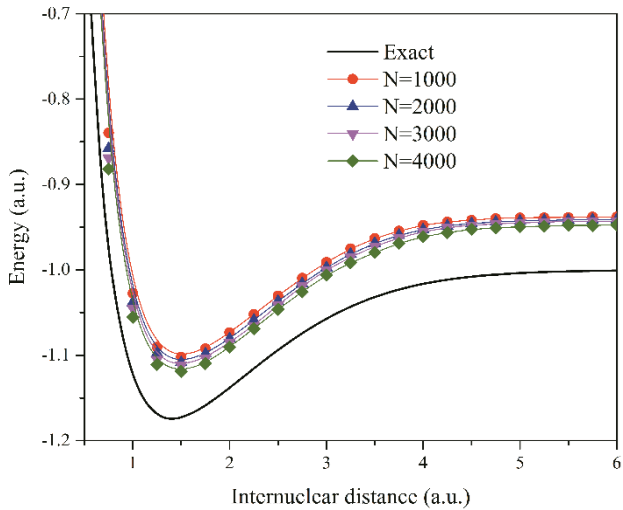

**FIG. 5.** (Color online) The potential well of H2 computed for different CS grids for the set of parameters ($\gamma = 0.75$, $\alpha_q = 10.0$, and $\alpha_p = 3.5$) and compared to the exact values (squares).

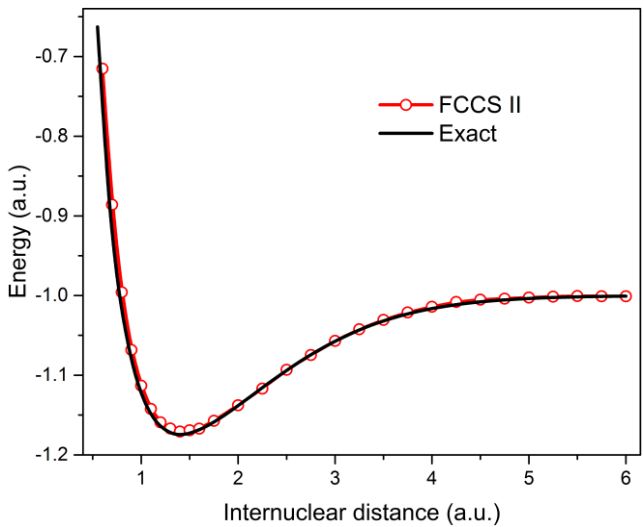

**FIG. 6.** (Color online) The potential well of H2 on the basis of FCCS-II computed for gird with N=1000 CS, compression parameters $\alpha_q = 10.0$ and $\alpha_p = 3.5$ and different gama parameters between 0.7-1.5 and compared with the exact potential [36-38]. The points of FCCS-II are shifted by same -0.064 a.u. value.

In FIG. 6, the potential well of $H_2$ on the basis of FCCS-II is computed for a grid with N=1000 CS, compression parameters $\alpha_q = 10.0$ and $\alpha_p = 3.5$ and compared with the exact curve. The energy of all points of the FCCS-II curve is shifted by -0.064 a.u. value. This figure represents that FCCS-II curve has a very good agreement with the exact curve especially for the equilibrium internuclear distance and dissociation energy. There are a slight different for the repulsion part of potential with respect of the exact.

## IV. CONCLUSIONS

One of the main goals in development of the CCS method is full dimensional study of many-electron systems such as He in the presence of an ultra-short intense laser field. Recently, FCCS-I, the modified version of the CCS method which is suitable for simulating fermionic systems has been introduced by Kirrander and Shalashilin [11]. There are some important problems in simulations on the basis of the CCS and the FCCS-I methods such as an essential grid refinement algorithm like the maximizing the residual overlap (MRO) in the DMC part and disability for incorporating more coupled CS into the simulation to approach more to the exact values. Moreover, implementation of the FCCS-I method due to its complex equation is cumbersome.

In this work, we introduced FCCS-II which does not have above-mentioned problems of CCS and FCCS-I. FCCS-II uses a simpler and more efficient approach for symmetrizing the spatial wave function in the simulation of fermionic systems in comparison with the approach used in FCCS-I. FCCS-II uses a self-symmetrized 6D CS grid and does not need any new equations other than the equations of the CCS method. All 6D CS in the grid constructed symmetrically with respect to the origin and indistinguishability of two electrons is implemented as explained in the Sec. II B. These advantages of FCCS-II significantly increase the speed of computations and give us capability to simulate the quantum systems with the larger CS grids. In addition, we considered two distinct compression parameters for position $\alpha_q$ and momentum $\alpha_p$. By this consideration, we can modify and adjust the CS grid more efficiently.

In summary, on the basis of FCCS-II, we applied the ITPSE method to gain the well potential of the ground state of H2 and the ground states of He. In comparison with the simulations having performed on the basis of the CCS method for gaining the electronic states of H2 which the numbers of the CS incorporating into the simulation was low (maximum 400 CS), we used more than 8000 CS in simulation of the ground state of He and the well potential of the ground state of H2. In all performed simulations, we looked for the best sets of parameters ($\gamma$, $\alpha_q$ and $\alpha_p$). For He, we found out that $\alpha_q = 1.5$ and $\alpha_p = 10.0$ are the best compression parameters. For computations based on small grids (< 200 CS), grids with 200 -1000 CS, and the larger grids (> 1000 CS), the best results corresponded to $\gamma \simeq 1.0$, 1.5, and 1.8, respectively. For a grid with 4000 CS, $\gamma \simeq 1.8$ leads to the lowest accessible energy. For $H_2$, for a grid with 500 CS, the lowest accessible energy obtained by $\gamma = 1.0, \alpha_q = 10.0, \alpha_p = 20.0$ and $\gamma = 0.75, \alpha_q = 10.0, \alpha_p = 20.0$ for small and large internuclear distances, respectively. For 1000 CS, the lowest accessible energy obtained by $\gamma = 1.0$ and $\gamma = 0.75$ for small and large internuclear distances, respectively.

In this work, we show that the FCCS-II with a relatively low number of CS (N=1000 CS) gives the potential energy curve for $H_2$ very close to the exact energy curve.

As a supplementary work, in this study, all the important equations of the FCCS-I method are re-derived. Furthermore, on the basis of FCCS-I, the dynamic equation of the wave function is proposed (Eqs. (39) And (40)).

What we have proposed in this article is a preliminary work to the simulations of two-electron systems in the presence of a high intense laser field, computing the high-order harmonic generation (HHG), and ionization rates in such systems on the basis of much larger CS grids.

## ACKNOWLEDGMENTS

M.E. and M.V. acknowledge partially financial support from Iran National Science Foundation: INSF with grant number 94003251. We want to convey our deepest appreciation to Prof. Dmitry Shalashilin for his generosity and helpful comments and suggestions that capable us to improve the manuscript more. M. E. would also like to thank Hamed Gholipour, Hossein Iravani, Hamed Ahmadi, Mitra Rooein and for reading this paper and giving useful comments.

## APENDIX A: ACCURATE DERIVATION OF THE MATRIX ELEMENTS OF THE LAGRANGIAN OPERATOR IN THE FCCS-I METHOD

In this Appendix, we derive the dynamic equations in the FCCS-I method in a new way. At first, let us derive the matrix elements of the Lagrangian operator Eq. (12) on the base of SCS Eq. (36)

$$\langle Z_s|\hat{L}|Z_s\rangle = \frac{i\hbar}{2}\left\langle Z_{e1}Z_{e2}\left|(2(1+a))^{-1/2}\frac{\vec{\partial}}{\partial t}(2(1+a))^{-1/2}\right|Z_{e1}Z_{e2}\right\rangle \quad (A1)$$
$$-\frac{i\hbar}{2}\left\langle Z_{e1}Z_{e2}\left|(2(1+a))^{-1/2}\frac{\overset{\leftarrow}{\partial}}{\partial t}(2(1+a))^{-1/2}\right|Z_{e1}Z_{e2}\right\rangle$$
$$+\frac{i\hbar}{2}\left\langle Z_{e1}Z_{e2}\left|(2(1+a))^{-1/2}\frac{\vec{\partial}}{\partial t}(2(1+a))^{-1/2}\right|Z_{e2}Z_{e1}\right\rangle$$
$$-\frac{i\hbar}{2}\left\langle Z_{e1}Z_{e2}\left|(2(1+a))^{-1/2}\frac{\overset{\leftarrow}{\partial}}{\partial t}(2(1+a))^{-1/2}\right|Z_{e2}Z_{e1}\right\rangle$$
$$+\frac{i\hbar}{2}\left\langle Z_{e2}Z_{e1}\left|(2(1+a))^{-1/2}\frac{\vec{\partial}}{\partial t}(2(1+a))^{-1/2}\right|Z_{e1}Z_{e2}\right\rangle$$
$$-\frac{i\hbar}{2}\left\langle Z_{e2}Z_{e1}\left|(2(1+a))^{-1/2}\frac{\overset{\leftarrow}{\partial}}{\partial t}(2(1+a))^{-1/2}\right|Z_{e1}Z_{e2}\right\rangle$$
$$+\frac{i\hbar}{2}\left\langle Z_{e2}Z_{e1}\left|(2(1+a))^{-1/2}\frac{\vec{\partial}}{\partial t}(2(1+a))^{-1/2}\right|Z_{e2}Z_{e1}\right\rangle$$
$$-\frac{i\hbar}{2}\left\langle Z_{e2}Z_{e1}\left|(2(1+a))^{-1/2}\frac{\overset{\leftarrow}{\partial}}{\partial t}(2(1+a))^{-1/2}\right|Z_{e2}Z_{e1}\right\rangle - \langle Z_s|\hat{H}|Z_s\rangle.$$

For the first term in the Eq. (A1) using Eqs. (20) and (22), we can write

$$\langle Z_{e1}|\dot{Z}_{e1}\rangle = \frac{1}{2}\sum_{j=1}^{M/2}\left(Z_j^*\dot{Z}_j - Z_j\dot{Z}_j^*\right) \quad (A2a)$$

$$\langle Z_{e2}|\dot{Z}_{e2}\rangle = \frac{1}{2}\sum_{j=1}^{M/2}\left(Z_{j+M/2}^*\dot{Z}_{j+M/2} - Z_{j+M/2}\dot{Z}_{j+M/2}^*\right) \quad (A2b)$$

$$\langle Z_{e1}|\dot{Z}_{e2}\rangle = \langle Z_{e1}|Z_{e2}\rangle\sum_{j=1}^{M/2}\left(Z_j^*\dot{Z}_{j+M/2}\right. \quad (A2c)$$
$$\left. -\frac{1}{2}\left(Z_{j+M/2}^*\dot{Z}_{j+M/2}\right.\right.$$
$$\left.\left. + Z_{j+M/2}\dot{Z}_{j+M/2}^*\right)\right)$$

$$\langle Z_{e2}|\dot{Z}_{e1}\rangle = \langle Z_{e2}|Z_{e1}\rangle \sum_{j=1}^{M/2} \left( Z^*_{j+M/2}\dot{Z}_j - \frac{1}{2}(Z^*_j\dot{Z}_j + Z_j\dot{Z}^*_j) \right) \quad \text{(A2d)}$$

$$\langle \dot{Z}_{e1}|Z_{e1}\rangle = \frac{1}{2}\sum_{j=1}^{M/2}(Z_j\,\dot{Z}^*_j - Z^*_j\dot{Z}_j) \quad \text{(A2e)}$$

$$\langle \dot{Z}_{e2}|Z_{e2}\rangle = \frac{1}{2}\sum_{j=1}^{M/2}(Z_{j+M/2}\dot{Z}^*_{j+M/2} - Z^*_{j+M/2}\dot{Z}_{j+M/2}) \quad \text{(A2f)}$$

$$\langle \dot{Z}_{e1}|Z_{e2}\rangle = \langle Z_{e1}|Z_{e2}\rangle \sum_{j=1}^{M/2}\left(Z_{j+M/2}\dot{Z}^*_j - \frac{1}{2}(Z_j\,\dot{Z}^*_j + Z^*_j\dot{Z}_j)\right) \quad \text{(A2g)}$$

$$\langle \dot{Z}_{e2}|Z_{e1}\rangle = \langle Z_{e2}|Z_{e1}\rangle \sum_{j=1}^{M/2}\Big(Z_j\dot{Z}^*_{j+M/2} - \frac{1}{2}(Z_{j+M/2}\,\dot{Z}^*_{j+M/2} + Z^*_{j+M/2}\dot{Z}_{j+M/2})\Big) \quad \text{(A2h)}$$

From Eq. (37) and Eq. (3) it can be seen that

$$a = \prod_{j=1}^{M/2} exp(Z^*_j Z_{j+M/2} + Z_j Z^*_{j+M/2} - Z_j Z^*_j - Z_{j+M/2} Z^*_{j+M/2}). \quad \text{(A3)}$$

Hence, related derivatives of above equation, can be achieved as follows

$$\frac{\partial a}{\partial Z_j} = (Z^*_{j+M/2} - Z^*_j)a \quad \text{(A4a)}$$

$$\frac{\partial a}{\partial Z_{j+M/2}} = (Z^*_j - Z^*_{j+M/2})a \quad \text{(A4b)}$$

$$\frac{\partial a}{\partial Z^*_j} = (Z_{j+M/2} - Z_j)a \quad \text{(A4c)}$$

$$\frac{\partial a}{\partial Z^*_{j+M/2}} = (Z_j - Z_{j+M/2})a. \quad \text{(A4d)}$$

Therefor for the first term in Eq. (A1) one can write

$$\left\langle Z_{e1}Z_{e2}\middle| (2(1+a))^{-1/2}\frac{\vec{\partial}}{\partial t}(2(1+a))^{-1/2}\middle| Z_{e1}Z_{e2}\right\rangle$$

$$= (2(1+a))^{-1/2}\left\langle Z_{e1}\middle|\frac{\vec{\partial}}{\partial t}(2(1+a))^{-1/2}\middle| Z_{e1}\right\rangle\langle Z_{e2}|Z_{e2}\rangle$$

$$+ (2(1+a))^{-1/2}\langle Z_{e1}|Z_{e1}\rangle\left\langle Z_{e2}\middle|\frac{\vec{\partial}}{\partial t}(2(1+a))^{-1/2}\middle| Z_{e2}\right\rangle$$

$$= \sum_{j=1}^{M/2}\left(\frac{1}{\sqrt{2(1+a)}}\frac{\partial}{\partial a}\left((2(1+a))^{-1/2}\right)\left(\frac{\partial a}{\partial Z_j}\frac{\partial Z_j}{\partial t} + \frac{\partial a}{\partial Z^*_j}\frac{\partial Z^*_j}{\partial t}\right)\langle Z_{e1}|Z_{e1}\rangle\langle Z_{e2}|Z_{e2}\rangle + \frac{1}{2(1+a)}\langle Z_{e1}|\dot{Z}_{e1}\rangle\langle Z_{e2}|Z_{e2}\rangle\right.$$

$$+ \frac{\langle Z_{e1}|Z_{e1}\rangle}{\sqrt{2(1+a)}}\frac{\partial}{\partial a}\left((2(1+a))^{-1/2}\right)\left(\frac{\partial a}{\partial Z_{j+M/2}}\frac{\partial Z_{j+M/2}}{\partial t} + \frac{\partial a}{\partial Z^*_{j+M/2}}\frac{\partial Z^*_{j+M/2}}{\partial t}\right)\langle Z_{e2}|Z_{e2}\rangle$$

$$\left.+ \frac{1}{2(1+a)}\langle Z_{e1}|Z_{e1}\rangle\langle Z_{e2}|\dot{Z}_{e2}\rangle\right).$$

Finally, the first term in Eq. (A1) has following form

$$\left\langle Z_{e1}Z_{e2}\middle| (2(1+a))^{-1/2}\frac{\vec{\partial}}{\partial t}(2(1+a))^{-1/2}\middle| Z_{e1}Z_{e2}\right\rangle \quad \text{(A5)}$$

$$= \sum_{j=1}^{M/2}\left(\left(\frac{1}{4(1+a)}Z^*_j - \frac{a}{(2(1+a))^2}(Z^*_{j+M/2} - Z^*_j)\right)\dot{Z}_j\right.$$

$$+ \left(\frac{1}{4(1+a)}Z^*_{j+M/2} - \frac{a}{(2(1+a))^2}(Z^*_j - Z^*_{j+M/2})\right)\dot{Z}_{j+M/2}$$

$$- \left(\frac{1}{4(1+a)}Z_j + \frac{a}{(2(1+a))^2}(Z_{j+M/2} - Z_j)\right)\dot{Z}^*_j$$

$$\left.- \left(\frac{1}{4(1+a)}Z_{j+M/2} + \frac{a}{(2(1+a))^2}(Z_j - Z_{j+M/2})\right)\dot{Z}^*_{j+M/2}\right).$$

For the seventh term in Eq. (A1), one can show that

$$\left\langle Z_{e2}Z_{e1} \left| (2(1+a))^{-1/2} \frac{\vec{\partial}}{\partial t} (2(1+a))^{-1/2} \right| Z_{e2}Z_{e1} \right\rangle = \left\langle Z_{e1}Z_{e2} \left| (2(1+a))^{-1/2} \frac{\vec{\partial}}{\partial t} (2(1+a))^{-1/2} \right| Z_{e1}Z_{e2} \right\rangle. \tag{A6}$$

Similarly, for other terms of the Eq. (A1), we have

$$\left\langle Z_{e1}Z_{e2} \left| (2(1+a))^{-1/2} \frac{\vec{\partial}}{\partial t} (2(1+a))^{-1/2} \right| Z_{e2}Z_{e1} \right\rangle \tag{A7}$$

$$= \sum_{j=1}^{M/2} \left( \left( \frac{a}{2(1+a)} \left( Z_j^* - \frac{1}{2} Z_{j+M/2}^* \right) - \frac{a^2}{(2(1+a))^2} \left( Z_j^* - Z_{j+M/2}^* \right) \right) \dot{Z}_{j+M/2} \right.$$

$$+ \left( \frac{a}{2(1+a)} \left( Z_{j+M/2}^* - \frac{1}{2} Z_j^* \right) - \frac{a^2}{(2(1+a))^2} \left( Z_{j+M/2}^* - Z_j^* \right) \right) \dot{Z}_j$$

$$- \left( \frac{a}{4(1+a)} Z_j + \frac{a^2}{(2(1+a))^2} \left( Z_{j+M/2} - Z_j \right) \right) \dot{Z}_j^*$$

$$\left. - \left( \frac{a}{4(1+a)} Z_{j+M/2} + \frac{a^2}{(2(1+a))^2} \left( Z_j - Z_{j+M/2} \right) \right) \dot{Z}_{j+M/2}^* \right)$$

$$\left\langle Z_{e2}Z_{e1} \left| (2(1+a))^{-1/2} \frac{\vec{\partial}}{\partial t} (2(1+a))^{-1/2} \right| Z_{e1}Z_{e2} \right\rangle = \left\langle Z_{e1}Z_{e2} \left| (2(1+a))^{-1/2} \frac{\vec{\partial}}{\partial t} (2(1+a))^{-1/2} \right| Z_{e2}Z_{e1} \right\rangle \tag{A8}$$

$$\left\langle Z_{e1}Z_{e2} \left| (2(1+a))^{-1/2} \frac{\vec{\partial}}{\partial t} (2(1+a))^{-1/2} \right| Z_{e1}Z_{e2} \right\rangle \tag{A9}$$

$$= \sum_{j=1}^{M/2} \left( \left( \frac{1}{4(1+a)} Z_j - \frac{a}{(2(1+a))^2} \left( Z_{j+M/2} - Z_j \right) \right) \dot{Z}_j^* \right.$$

$$+ \left( \frac{1}{4(1+a)} Z_{j+M/2} - \frac{a}{(2(1+a))^2} \left( Z_j - Z_{j+M/2} \right) \right) \dot{Z}_{j+M/2}^*$$

$$- \left( \frac{1}{4(1+a)} Z_j^* + \frac{a}{(2(1+a))^2} \left( Z_{j+M/2}^* - Z_j^* \right) \right) \dot{Z}_j$$

$$\left. - \left( \frac{1}{4(1+a)} Z_{j+M/2}^* + \frac{a}{(2(1+a))^2} \left( Z_j^* - Z_{j+M/2}^* \right) \right) \dot{Z}_{j+M/2} \right)$$

$$\left\langle Z_{e2}Z_{e1} \left| (2(1+a))^{-1/2} \frac{\vec{\partial}}{\partial t} (2(1+a))^{-1/2} \right| Z_{e2}Z_{e1} \right\rangle = \left\langle Z_{e1}Z_{e2} \left| (2(1+a))^{-1/2} \frac{\vec{\partial}}{\partial t} (2(1+a))^{-1/2} \right| Z_{e1}Z_{e2} \right\rangle \tag{A10}$$

$$\left\langle Z_{e1}Z_{e2} \left| (2(1+a))^{-1/2} \frac{\vec{\partial}}{\partial t} (2(1+a))^{-1/2} \right| Z_{e2}Z_{e1} \right\rangle \tag{A11}$$

$$= \sum_{j=1}^{M/2} \left( \left( \frac{a \left( Z_{j+M/2} - \frac{1}{2} Z_j \right)}{2(1+a)} - \frac{a^2}{(2(1+a))^2} \left( Z_{j+M/2} - Z_j \right) \right) \dot{Z}_j^* \right.$$

$$+ \left( \frac{a \left( Z_j - \frac{1}{2} Z_{j+M/2} \right)}{2(1+a)} - \frac{a^2}{(2(1+a))^2} \left( Z_j - Z_{j+M/2} \right) \right) \dot{Z}_{j+M/2}^*$$

$$- \left( \frac{a}{4(1+a)} Z_j^* + \frac{a^2}{(2(1+a))^2} \left( Z_{j+M/2}^* - Z_j^* \right) \right) \dot{Z}_j$$

$$\left. - \left( \frac{a}{4(1+a)} Z_{j+M/2}^* + \frac{a^2}{(2(1+a))^2} \left( Z_j^* - Z_{j+M/2}^* \right) \right) \dot{Z}_{j+M/2} \right)$$

$$\left\langle Z_{e2}Z_{e1}\left|(2(1+a))^{-1/2}\frac{\vec{\partial}}{\partial t}(2(1+a))^{-1/2}\right|Z_{e1}Z_{e2}\right\rangle = \left\langle Z_{e1}Z_{e2}\left|(2(1+a))^{-1/2}\frac{\vec{\partial}}{\partial t}(2(1+a))^{-1/2}\right|Z_{e2}Z_{e1}\right\rangle. \tag{A12}$$

At the end, by substituting Eqs. (A6) – (A12) in Eq. (A1) and doing some straightforward simplifications, one can derive the matrix element of the Lagarangian on the base of SCS

$$\ell = \langle Z_s|\hat{\mathcal{L}}|Z_s\rangle = \frac{i\hbar}{2(1+a)}\sum_{j=1}^{M/2}\left((Z_j^* + aZ_{j+M/2}^*)\dot{Z}_j + (Z_{j+M/2}^* + aZ_j^*)\dot{Z}_{j+M/2} - (Z_j + aZ_{j+M/2})\dot{Z}_j^* - (Z_{j+M/2} + aZ_j^*)\dot{Z}_{j+M/2}^*\right) - \langle Z_s|\hat{H}|Z_s\rangle. \tag{A13}$$

In the Eq. (A13), the time derivatives of the eigenvalues of coherent states, can be computed as follows

$$\dot{Z}_{j,j+M/2} = -\frac{i}{\hbar}\frac{\partial\langle Z_s|\hat{H}|Z_s\rangle}{\partial Z_{j,j+M/2}^*} \tag{A14}$$

$$\dot{Z}_{j,j+M/2}^* = \frac{i}{\hbar}\frac{\partial\langle Z_s|\hat{H}|Z_s\rangle}{\partial Z_{j,j+M/2}}$$

where on the base of SCS, the matrix elements of the Hamiltonian of the system is

$$\langle Z_s|\hat{H}|Z_s\rangle = \frac{1}{2(1+a)}\left[\langle Z_{e1}Z_{e2}|\hat{H}|Z_{e1}Z_{e2}\rangle + \langle Z_{e1}Z_{e2}|\hat{H}|Z_{e2}Z_{e1}\rangle + \langle Z_{e2}Z_{e1}|\hat{H}|Z_{e1}Z_{e2}\rangle + \langle Z_{e2}Z_{e1}|\hat{H}|Z_{e2}Z_{e1}\rangle\right]. \tag{A15}$$

Using Eq. (4), it can be seen that

$$\langle Z_s|\hat{H}|Z_s\rangle = \frac{1}{2(1+a)}[\widetilde{H}(Z_{12}^*,Z_{12})\langle Z_{e1}Z_{e2}|Z_{e1}Z_{e2}\rangle + \widetilde{H}(Z_{12}^*,Z_{21})\langle Z_{e1}Z_{e2}|Z_{e2}Z_{e1}\rangle + \widetilde{H}(Z_{21}^*,Z_{12})\langle Z_{e2}Z_{e1}|Z_{e1}Z_{e2}\rangle + \widetilde{H}(Z_{21}^*,Z_{21})\langle Z_{e2}Z_{e1}|Z_{e2}Z_{e1}\rangle]. \tag{A16}$$

One can show that

$$\langle Z_{e1}Z_{e2}|Z_{e1}Z_{e2}\rangle = \langle Z_{e2}Z_{e1}|Z_{e2}Z_{e1}\rangle = 1 \tag{A17a}$$

$$\widetilde{H}(Z_{12}^*,Z_{12}) = \widetilde{H}(Z_{21}^*,Z_{21}) \tag{A17b}$$

$$\langle Z_{e1}Z_{e2}|Z_{e2}Z_{e1}\rangle = \langle Z_{e2}Z_{e1}|Z_{e1}Z_{e2}\rangle = a \tag{A17c}$$

$$\widetilde{H}(Z_{12}^*,Z_{21}) = \widetilde{H}(Z_{21}^*,Z_{12}). \tag{A17d}$$

Hence, the matrix elements of the Hamiltonian of the system on the base of SCS have this form

$$\langle Z_s|\hat{H}|Z_s\rangle = \langle\hat{H}\rangle = \frac{\widetilde{H}(Z_{12}^*,Z_{12}) + a\widetilde{H}(Z_{12}^*,Z_{21})}{1+a}. \tag{A18}$$

Using Eq. (18), for the time derivatives of the eigenvalues of coherent states Eq. (A14), we can write

$$\frac{\partial\langle\hat{H}\rangle}{\partial Z_{j,j+M/2}^*} = \widetilde{H}(Z_{12}^*,Z_{12})\frac{\partial}{\partial Z_{j,j+M/2}^*}\left(\frac{1}{1+a}\right) + \frac{1}{1+a}\frac{\partial\widetilde{H}(Z_{12}^*,Z_{12})}{\partial Z_{j,j+M/2}^*} + \widetilde{H}(Z_{12}^*,Z_{21})\frac{\partial}{\partial Z_{j,j+M/2}^*}\left(\frac{a}{1+a}\right) + \frac{a}{1+a}\frac{\partial\widetilde{H}(Z_{12}^*,Z_{21})}{\partial Z_{j,j+M/2}^*} \tag{A19a}$$

$$\frac{\partial\langle\hat{H}\rangle}{\partial Z_{j,j+M/2}} = \widetilde{H}(Z_{12}^*,Z_{12})\frac{\partial}{\partial Z_{j,j+M/2}}\left(\frac{1}{1+a}\right) + \frac{1}{1+a}\frac{\partial\widetilde{H}(Z_{12}^*,Z_{12})}{\partial Z_{j,j+M/2}} + \widetilde{H}(Z_{12}^*,Z_{21})\frac{\partial}{\partial Z_{j,j+M/2}}\left(\frac{a}{1+a}\right) + \frac{a}{1+a}\frac{\partial\widetilde{H}(Z_{12}^*,Z_{21})}{\partial Z_{j,j+M/2}} \tag{A19b}$$

where

$$\frac{\partial}{\partial Z_j^*}\left(\frac{1}{1+a}\right) = \frac{\partial}{\partial a}\left(\frac{1}{1+a}\right)\frac{\partial a}{\partial Z_j^*} = \frac{-a}{(1+a)^2}(Z_{j+M/2} - Z_j) \tag{A20a}$$

$$\frac{\partial}{\partial Z_j}\left(\frac{1}{1+a}\right) = \frac{\partial}{\partial a}\left(\frac{1}{1+a}\right)\frac{\partial a}{\partial Z_j} = \frac{-a}{(1+a)^2}(Z_{j+M/2}^* - Z_j^*) \tag{A20b}$$

$$\frac{\partial}{\partial Z_{j+M/2}^*}\left(\frac{a}{1+a}\right) = \frac{\partial}{\partial a}\left(\frac{a}{1+a}\right)\frac{\partial a}{\partial Z_{j+M/2}^*} = \frac{a}{(1+a)^2}(Z_j - Z_{j+M/2}) \tag{A20c}$$

$$\frac{\partial}{\partial Z_{j+M/2}}\left(\frac{a}{1+a}\right) = \frac{\partial}{\partial a}\left(\frac{a}{1+a}\right)\frac{\partial a}{\partial Z_{j+M/2}} = \frac{a}{(1+a)^2}(Z_j^* - Z_{j+M/2}^*). \tag{A20d}$$

Hence

$$\dot{Z}_j = -\frac{i}{\hbar}\frac{\partial \langle \widehat{H}\rangle}{\partial Z_j^*} = -\frac{i}{\hbar}\left(\frac{-a}{(1+a)^2}(Z_{j+M/2}-Z_j)\widetilde{H}(Z_{12}^*,Z_{12}) + \frac{1}{1+a}\frac{\partial \widetilde{H}(Z_{12}^*,Z_{12})}{\partial Z_j^*} + \frac{a}{(1+a)^2}(Z_{j+M/2}-Z_j)\widetilde{H}(Z_{12}^*,Z_{21})\right.$$
$$\left.+ \frac{a}{1+a}\frac{\partial \widetilde{H}(Z_{12}^*,Z_{21})}{\partial Z_j^*}\right) \tag{A21a}$$

$$\dot{Z}_{j+M/2} = -\frac{i}{\hbar}\frac{\partial \langle \widehat{H}\rangle}{\partial Z_{j+M/2}^*} \tag{A21b}$$
$$= -\frac{i}{\hbar}\left(\frac{-a}{(1+a)^2}(Z_j - Z_{j+M/2})\widetilde{H}(Z_{12}^*,Z_{12}) + \frac{1}{1+a}\frac{\partial \widetilde{H}(Z_{12}^*,Z_{12})}{\partial Z_{j+M/2}^*} + \frac{a}{(1+a)^2}(Z_j - Z_{j+M/2})\widetilde{H}(Z_{12}^*,Z_{21})\right.$$
$$\left.+ \frac{a}{1+a}\frac{\partial \widetilde{H}(Z_{12}^*,Z_{21})}{\partial Z_{j+M/2}^*}\right)$$

$$\dot{Z}_j^* = \frac{i}{\hbar}\frac{\partial \langle \widehat{H}\rangle}{\partial Z_j} = \frac{i}{\hbar}\left(\frac{-a}{(1+a)^2}(Z_{j+M/2}^* - Z_j^*)\widetilde{H}(Z_{12}^*,Z_{12}) + \frac{1}{1+a}\frac{\partial \widetilde{H}(Z_{12}^*,Z_{12})}{\partial Z_j} + \frac{a}{(1+a)^2}(Z_{j+M/2}^* - Z_j^*)\widetilde{H}(Z_{12}^*,Z_{21})\right.$$
$$\left.+ \frac{a}{1+a}\frac{\partial \widetilde{H}(Z_{12}^*,Z_{21})}{\partial Z_j}\right) \tag{A22c}$$

$$\dot{Z}_{j+M/2}^* = \frac{i}{\hbar}\frac{\partial \langle \widehat{H}\rangle}{\partial Z_{j+M/2}} \tag{A22d}$$
$$= \frac{i}{\hbar}\left(\frac{-a}{(1+a)^2}(Z_j^* - Z_{j+M/2}^*)\widetilde{H}(Z_{12}^*,Z_{12}) + \frac{1}{1+a}\frac{\partial \widetilde{H}(Z_{12}^*,Z_{12})}{\partial Z_{j+M/2}} + \frac{a}{(1+a)^2}(Z_j^* - Z_{j+M/2}^*)\widetilde{H}(Z_{12}^*,Z_{21})\right.$$
$$\left.+ \frac{a}{1+a}\frac{\partial \widetilde{H}(Z_{12}^*,Z_{21})}{\partial Z_{j+M/2}}\right)$$

## APPENDIX B: THE MATRIX ELEMENTS OF THE HAMILTONIAN AND ITS DERIVATIVES IN THE FCCS-I METHOD

On the base of SCS (Eq. (36)), the first part of the matrix elements of the Hamiltonian $\widetilde{H}(Z_{12}^*, Z_{12})$ in Eq. (A18) and its derivatives can be derived as follows

$$\widetilde{H}(Z_{12}^*, Z_{12}) = \frac{P_{e1}^2(Z_{12}^*, Z_{12})}{2} + \frac{P_{e2}^2(Z_{12}^*, Z_{12})}{2} - \frac{1}{\sqrt{\boldsymbol{\rho}_{1(1)}^{(11)^2}(Z_{12}^*, Z_{12})}} erf\left(\sqrt{\gamma \boldsymbol{\rho}_{1(1)}^{(11)^2}(Z_{12}^*, Z_{12})}\right) \tag{B1}$$
$$- \frac{1}{\sqrt{\boldsymbol{\rho}_{1(2)}^{(11)^2}(Z_{12}^*, Z_{12})}} erf\left(\sqrt{\gamma \boldsymbol{\rho}_{1(2)}^{(11)^2}(Z_{12}^*, Z_{12})}\right) - \frac{1}{\sqrt{\boldsymbol{\rho}_{2(1)}^{(11)^2}(Z_{12}^*, Z_{12})}} erf\left(\sqrt{\gamma \boldsymbol{\rho}_{2(1)}^{(11)^2}(Z_{12}^*, Z_{12})}\right)$$
$$- \frac{1}{\sqrt{\boldsymbol{\rho}_{2(2)}^{(11)^2}(Z_{12}^*, Z_{12})}} erf\left(\sqrt{\gamma \boldsymbol{\rho}_{2(2)}^{(11)^2}(Z_{12}^*, Z_{12})}\right) + \frac{1}{\sqrt{\boldsymbol{\rho}_{12}^{(11)^2}(Z_{12}^*, Z_{12})}} erf\left(\sqrt{\frac{\gamma}{2}\boldsymbol{\rho}_{12}^{(11)^2}(Z_{12}^*, Z_{12})}\right),$$

where in Eq. (B1) for the first two terms and their derivatives we have

$$P_{e1}^2(Z_{12}^*, Z_{12}) = -\frac{\gamma}{2}\sum_{j=1}^{M/2}\left(Z_j^{*2} + Z_j^2 - 2Z_j^*Z_j - 1\right) \tag{B2a}$$

$$P_{e2}^2(Z_{12}^*, Z_{12}) = -\frac{\gamma}{2}\sum_{j=1}^{M/2}\left(Z_{j+M/2}^{*^2} + Z_{j+M/2}^2 - 2Z_{j+M/2}^*Z_{j+M/2} - 1\right) \tag{B2b}$$

$$\frac{\partial\left(P_{e1}^2(Z_{12}^*, Z_{12}) + P_{e2}^2(Z_{12}^*, Z_{12})\right)}{\partial Z_{j,j+M/2}^*} = -\gamma\left(Z_{j,j+M/2}^* - Z_{j,j+M/2}\right) \tag{B2c}$$

$$\frac{\partial\left(P_{e1}^2(Z_{12}^*, Z_{12}) + P_{e2}^2(Z_{12}^*, Z_{12})\right)}{\partial Z_{j,j+M/2}} = -\gamma\left(Z_{j,j+M/2} - Z_{j,j+M/2}^*\right). \tag{B2d}$$

For the Coulombic potentials in Eq. (B1) and their related derivatives, one can verify that

$$\boldsymbol{\rho}_{1(k)}^{(11)}(Z_{12}^*, Z_{12}) = \frac{Z_{e1}^* + Z_{e1}}{\sqrt{2\gamma}} - \boldsymbol{R}_k \qquad k = 1,2 \tag{B3a}$$

$$\frac{\partial \boldsymbol{\rho}_{1(k)}^{(11)^2}(Z_{12}^*,Z_{12})}{\partial Z_j^*} = \frac{\partial}{\partial Z_j^*}\sum_{j=1}^{M/2}\left(\frac{Z_j^*+Z_j}{\sqrt{2\gamma}}-\boldsymbol{R}_k\right)^2 = \sqrt{\frac{2}{\gamma}}\left(\frac{Z_j^*+Z_j}{\sqrt{2\gamma}}-\boldsymbol{R}_k\right) \quad k=1,2 \tag{B3b}$$

$$\frac{\partial \boldsymbol{\rho}_{1(k)}^{(11)^2}(Z_{12}^*,Z_{12})}{\partial Z_{j+M/2}^*} = 0 \quad k=1,2 \tag{B3c}$$

$$\frac{\partial \boldsymbol{\rho}_{1(k)}^{(11)^2}(Z_{12}^*,Z_{12})}{\partial Z_j} = \frac{\partial}{\partial Z_j}\sum_{j=1}^{M/2}\left(\frac{Z_j^*+Z_j}{\sqrt{2\gamma}}-\boldsymbol{R}_k\right)^2 = \sqrt{\frac{2}{\gamma}}\left(\frac{Z_j^*+Z_j}{\sqrt{2\gamma}}-\boldsymbol{R}_k\right) \quad k=1,2 \tag{B3d}$$

$$\frac{\partial \boldsymbol{\rho}_{1(k)}^{(11)^2}(Z_{12}^*,Z_{12})}{\partial Z_{j+M/2}} = 0 \quad k=1,2 \tag{B3e}$$

$$\boldsymbol{\rho}_{2(k)}^{(11)}(Z_{12}^*,Z_{12}) = \frac{Z_{e2}^*+Z_{e2}}{\sqrt{2\gamma}}-\boldsymbol{R}_k \quad k=1,2 \tag{B3f}$$

$$\frac{\partial \boldsymbol{\rho}_{2(k)}^{(11)^2}(Z_{12}^*,Z_{12})}{\partial Z_j^*} = 0 \quad k=1,2 \tag{B3g}$$

$$\frac{\partial \boldsymbol{\rho}_{2(k)}^{(11)^2}(Z_{12}^*,Z_{12})}{\partial Z_{j+M/2}^*} = \frac{\partial}{\partial Z_{j+M/2}^*}\sum_{j=1}^{M/2}\left(\frac{Z_{j+M/2}^*+Z_{j+M/2}}{\sqrt{2\gamma}}-\boldsymbol{R}_k\right)^2 = \sqrt{\frac{2}{\gamma}}\left(\frac{Z_{j+M/2}^*+Z_{j+M/2}}{\sqrt{2\gamma}}-\boldsymbol{R}_k\right) \quad k=1,2 \tag{B3h}$$

$$\frac{\partial \boldsymbol{\rho}_{2(k)}^{(11)^2}(Z_{12}^*,Z_{12})}{\partial Z_j} = 0 \quad k=1,2 \tag{B3i}$$

$$\frac{\partial \boldsymbol{\rho}_{2(k)}^{(11)^2}(Z_{12}^*,Z_{12})}{\partial Z_{j+M/2}} = \frac{\partial}{\partial Z_{j+M/2}}\sum_{j=1}^{M/2}\left(\frac{Z_{j+M/2}^*+Z_{j+M/2}}{\sqrt{2\gamma}}-\boldsymbol{R}_k\right)^2 = \sqrt{\frac{2}{\gamma}}\left(\frac{Z_{j+M/2}^*+Z_{j+M/2}}{\sqrt{2\gamma}}-\boldsymbol{R}_k\right) \quad k=1,2 \tag{B3j}$$

$$\boldsymbol{\rho}_{12}^{(11)}(Z_{12}^*,Z_{12}) = \frac{Z_{e1}^*+Z_{e1}}{\sqrt{2\gamma}} - \frac{Z_{e2}^*+Z_{e2}'}{\sqrt{2\gamma}} \tag{B3k}$$

$$\frac{\partial \boldsymbol{\rho}_{12}^{(11)^2}(Z_{12}^*,Z_{12})}{\partial Z_j^*} = \frac{\partial}{\partial Z_j^*}\sum_{j=1}^{M/2}\left(\frac{Z_j^*+Z_j}{\sqrt{2\gamma}}-\frac{Z_{j+M/2}^*+Z_{j+M/2}}{\sqrt{2\gamma}}\right)^2 = \frac{1}{\gamma}\left(Z_j^*+Z_j-Z_{j+M/2}^*-Z_{j+M/2}\right) \tag{B3l}$$
$$= -\frac{\partial \boldsymbol{\rho}_{12}^{(11)^2}(Z_{12}^*,Z_{12})}{\partial Z_{j+M/2}^*} \quad k=1,2$$

$$\frac{\partial \boldsymbol{\rho}_{12}^{(11)^2}(Z_{12}^*,Z_{12})}{\partial Z_j} = \frac{\partial}{\partial Z_j}\sum_{j=1}^{M/2}\left(\frac{Z_j^*+Z_j}{\sqrt{2\gamma}}-\frac{Z_{j+M/2}^*+Z_{j+M/2}}{\sqrt{2\gamma}}\right)^2 = \frac{1}{\gamma}\left(Z_j^*+Z_j-Z_{j+M/2}^*-Z_{j+M/2}\right) \tag{B3m}$$
$$= -\frac{\partial \boldsymbol{\rho}_{12}^{(11)^2}(Z_{12}^*,Z_{12})}{\partial Z_{j+M/2}} \quad k=1,2 \; .$$

Therefor $Z^*$ derivatives of $\widetilde{H}(Z_{12}^*,Z_{12})$ can be derived as follows

$$\frac{\partial \widetilde{H}(Z^*_{12}, Z_{12})}{\partial Z^*_{j,j+M/2}} = -\frac{\gamma}{2}\left(Z^*_{j,j+M/2} - Z_{j,j+M/2}\right) \tag{B4}$$

$$-\frac{\partial \rho^{(11)^2}_{1(1)}(Z^*_{12}, Z_{12})}{\partial Z^*_{j,j+M/2}}\left(\frac{-erf\left(\sqrt{\gamma \rho^{(11)^2}_{1(1)}(Z^*_{12}, Z_{12})}\right)}{2\sqrt{\rho^{(11)}_{1(1)}(Z^*_{12}, Z_{12})}^3} + \sqrt{\frac{\gamma}{\pi}}\frac{exp\left(-\gamma\sqrt{\rho^{(11)}_{1(1)}(Z^*_{12}, Z_{12})}^2\right)}{\sqrt{\rho^{(11)}_{1(1)}(Z^*_{12}, Z_{12})}^2}\right)$$

$$-\frac{\partial \rho^{(11)^2}_{1(2)}(Z^*_{12}, Z_{12})}{\partial Z^*_{j,j+M/2}}\left(\frac{-erf\left(\sqrt{\gamma \rho^{(11)^2}_{1(2)}(Z^*_{12}, Z_{12})}\right)}{2\sqrt{\rho^{(11)}_{1(2)}(Z^*_{12}, Z_{12})}^3} + \sqrt{\frac{\gamma}{\pi}}\frac{exp\left(-\gamma\sqrt{\rho^{(11)}_{1(2)}(Z^*_{12}, Z_{12})}^2\right)}{\sqrt{\rho^{(11)}_{1(2)}(Z^*_{12}, Z_{12})}^2}\right)$$

$$-\frac{\partial \rho^{(11)^2}_{2(1)}(Z^*_{12}, Z_{12})}{\partial Z^*_{j,j+M/2}}\left(\frac{-erf\left(\sqrt{\gamma \rho^{(11)^2}_{2(1)}(Z^*_{12}, Z_{12})}\right)}{2\sqrt{\rho^{(11)}_{2(1)}(Z^*_{12}, Z_{12})}^3} + \sqrt{\frac{\gamma}{\pi}}\frac{exp\left(-\gamma\sqrt{\rho^{(11)}_{2(1)}(Z^*_{12}, Z_{12})}^2\right)}{\sqrt{\rho^{(11)}_{2(1)}(Z^*_{12}, Z_{12})}^2}\right)$$

$$-\frac{\partial \rho^{(11)^2}_{2(2)}(Z^*_{12}, Z_{12})}{\partial Z^*_{j,j+M/2}}\left(\frac{-erf\left(\sqrt{\gamma \rho^{(11)^2}_{2(2)}(Z^*_{12}, Z_{12})}\right)}{2\sqrt{\rho^{(11)}_{2(2)}(Z^*_{12}, Z_{12})}^3} + \sqrt{\frac{\gamma}{\pi}}\frac{exp\left(-\gamma\sqrt{\rho^{(11)}_{2(2)}(Z^*_{12}, Z_{12})}^2\right)}{\sqrt{\rho^{(11)}_{2(2)}(Z^*_{12}, Z_{12})}^2}\right)$$

$$+\frac{\partial \rho^{(11)^2}_{12}(Z^*_{12}, Z_{12})}{\partial Z^*_{j,j+M/2}}\left(\frac{-erf\left(\sqrt{\frac{\gamma}{2} \rho^{(11)^2}_{12}(Z^*_{12}, Z_{12})}\right)}{2\sqrt{\rho^{(11)}_{12}(Z^*_{12}, Z_{12})}^3} + \sqrt{\frac{\gamma}{2\pi}}\frac{exp\left(-\frac{\gamma}{2}\sqrt{\rho^{(11)}_{12}(Z^*_{12}, Z_{12})}^2\right)}{\sqrt{\rho^{(11)}_{12}(Z^*_{12}, Z_{12})}^2}\right)$$

and similarly, for $Z$ derivatives of $\widetilde{H}(Z^*_{12}, Z_{12})$ we would have

$$\frac{\partial \widetilde{H}(Z^*_{12}, Z_{12})}{\partial Z_{j,j+M/2}} = -\gamma\left(Z_{j,j+M/2} - Z^*_{j,j+M/2}\right) \tag{B5}$$

$$-\frac{\partial \rho^{(11)^2}_{1(1)}(Z^*_{12}, Z_{12})}{\partial Z_{j,j+M/2}}\left(\frac{-erf\left(\sqrt{\gamma \rho^{(11)^2}_{1(1)}(Z^*_{12}, Z_{12})}\right)}{2\sqrt{\rho^{(11)}_{1(1)}(Z^*_{12}, Z_{12})}^3} + \sqrt{\frac{\gamma}{\pi}}\frac{exp\left(-\gamma\sqrt{\rho^{(11)}_{1(1)}(Z^*_{12}, Z_{12})}^2\right)}{\sqrt{\rho^{(11)}_{1(1)}(Z^*_{12}, Z_{12})}^2}\right)$$

$$-\frac{\partial \rho^{(11)^2}_{1(2)}(Z^*_{12}, Z_{12})}{\partial Z_{j,j+M/2}}\left(\frac{-erf\left(\sqrt{\gamma \rho^{(11)^2}_{1(2)}(Z^*_{12}, Z_{12})}\right)}{2\sqrt{\rho^{(11)}_{1(2)}(Z^*_{12}, Z_{12})}^3} + \sqrt{\frac{\gamma}{\pi}}\frac{exp\left(-\gamma\sqrt{\rho^{(11)}_{1(2)}(Z^*_{12}, Z_{12})}^2\right)}{\sqrt{\rho^{(11)}_{1(2)}(Z^*_{12}, Z_{12})}^2}\right)$$

$$-\frac{\partial \rho^{(11)^2}_{2(1)}(Z^*_{12}, Z_{12})}{\partial Z_{j,j+M/2}}\left(\frac{-erf\left(\sqrt{\gamma \rho^{(11)^2}_{2(1)}(Z^*_{12}, Z_{12})}\right)}{2\sqrt{\rho^{(11)}_{2(1)}(Z^*_{12}, Z_{12})}^3} + \sqrt{\frac{\gamma}{\pi}}\frac{exp\left(-\gamma\sqrt{\rho^{(11)}_{2(1)}(Z^*_{12}, Z_{12})}^2\right)}{\sqrt{\rho^{(11)}_{2(1)}(Z^*_{12}, Z_{12})}^2}\right)$$

$$-\frac{\partial \rho^{(11)^2}_{2(2)}(Z^*_{12}, Z_{12})}{\partial Z_{j,j+M/2}}\left(\frac{-erf\left(\sqrt{\gamma \rho^{(11)^2}_{2(2)}(Z^*_{12}, Z_{12})}\right)}{2\sqrt{\rho^{(11)}_{2(2)}(Z^*_{12}, Z_{12})}^3} + \sqrt{\frac{\gamma}{\pi}}\frac{exp\left(-\gamma\sqrt{\rho^{(11)}_{2(2)}(Z^*_{12}, Z_{12})}^2\right)}{\sqrt{\rho^{(11)}_{2(2)}(Z^*_{12}, Z_{12})}^2}\right)$$

$$+\frac{\partial \rho^{(11)^2}_{12}(Z^*_{12}, Z_{12})}{\partial Z_{j,j+M/2}}\left(\frac{-erf\left(\sqrt{\frac{\gamma}{2} \rho^{(11)^2}_{12}(Z^*_{12}, Z_{12})}\right)}{2\sqrt{\rho^{(11)}_{12}(Z^*_{12}, Z_{12})}^3} + \sqrt{\frac{\gamma}{2\pi}}\frac{exp\left(-\frac{\gamma}{2}\sqrt{\rho^{(11)}_{12}(Z^*_{12}, Z_{12})}^2\right)}{\sqrt{\rho^{(11)}_{12}(Z^*_{12}, Z_{12})}^2}\right).$$

On the base of SCS Eq. (36), for the second part of the matrix elements of the Hamiltonian $\widetilde{H}(Z^*_{12}, Z_{21})$ in Eq. (A18) and its derivatives we can write

$$\widetilde{H}(Z_{12}^*, Z_{21}) = \frac{P_{e1}^2(Z_{12}^*, Z_{21})}{2} + \frac{P_{e2}^2(Z_{12}^*, Z_{21})}{2} - \frac{1}{\left|\boldsymbol{\rho}_{1(1)}^{(12)}(Z_{12}^*, Z_{21})\right|} erf\left(\sqrt{\gamma}\left|\boldsymbol{\rho}_{1(1)}^{(12)}(Z_{12}^*, Z_{21})\right|\right) \quad \text{(B6)}$$

$$- \frac{1}{\left|\boldsymbol{\rho}_{1(2)}^{(12)}(Z_{12}^*, Z_{21})\right|} erf\left(\sqrt{\gamma}\left|\boldsymbol{\rho}_{1(2)}^{(12)}(Z_{12}^*, Z_{21})\right|\right) - \frac{1}{\left|\boldsymbol{\rho}_{2(1)}^{(12)}(Z_{12}^*, Z_{21})\right|} erf\left(\sqrt{\gamma}\left|\boldsymbol{\rho}_{2(1)}^{(12)}(Z_{12}^*, Z_{21})\right|\right)$$

$$- \frac{1}{\left|\boldsymbol{\rho}_{2(2)}^{(12)}(Z_{12}^*, Z_{21})\right|} erf\left(\sqrt{\gamma}\left|\boldsymbol{\rho}_{2(2)}^{(12)}(Z_{12}^*, Z_{21})\right|\right) + \frac{1}{\left|\boldsymbol{\rho}_{12}^{(12)}(Z_{12}^*, Z_{21})\right|} erf\left(\sqrt{\gamma/2}\left|\boldsymbol{\rho}_{12}^{(12)}(Z_{12}^*, Z_{21})\right|\right)$$

$$+ \frac{1}{|\boldsymbol{R}_1 - \boldsymbol{R}_2|}$$

where in above equation for the first two terms and their derivatives we have

$$P_{e1}^2(Z_{12}^*, Z_{21}) = -\frac{\gamma}{2}\sum_{j=1}^{M/2}\left(Z_j^{*2} + Z_{j+M/2}^2 - 2Z_j^* Z_{j+M/2} - 1\right) \quad \text{(B7a)}$$

$$P_{e2}^2(Z_{12}^*, Z_{21}) = -\frac{\gamma}{2}\sum_{j=1}^{M/2}\left({Z_{j+M/2}^*}^2 + Z_j^2 - 2Z_{j+M/2}^* Z_j - 1\right) \quad \text{(B7b)}$$

$$\frac{\partial\left(P_{e1}^2(Z_{12}^*, Z_{21}) + P_{e2}^2(Z_{12}^*, Z_{21})\right)}{\partial Z_{j,j+M/2}^*} = -\gamma\left(Z_{j,j+M/2}^* - Z_{j+M/2,j}\right) \quad \text{(B7c)}$$

$$\frac{\partial\left(P_{e1}^2(Z_{12}^*, Z_{21}) + P_{e2}^2(Z_{12}^*, Z_{21})\right)}{\partial Z_{j,j+M/2}} = -\gamma\left(Z_{j,j+M/2} - Z_{j+M/2,j}^*\right). \quad \text{(B7d)}$$

For the Coulombic potentials in Eq. (B6) and their related derivatives, one can verify that

$$\boldsymbol{\rho}_{1(k)}^{(12)}(Z_{12}^*, Z_{21}) = \frac{Z_{e1}^* + Z_{e2}}{\sqrt{2\gamma}} - \boldsymbol{R}_k \qquad k = 1,2 \quad \text{(B8a)}$$

$$\frac{\partial {\boldsymbol{\rho}_{1(k)}^{(12)}}^2(Z_{12}^*, Z_{21})}{\partial Z_j^*} = \frac{\partial}{\partial Z_j^*}\sum_{j=1}^{M/2}\left(\frac{Z_j^* + Z_{j+M/2}}{\sqrt{2\gamma}} - \boldsymbol{R}_k\right)^2 = \sqrt{\frac{2}{\gamma}}\left(\frac{Z_j^* + Z_{j+M/2}}{\sqrt{2\gamma}} - \boldsymbol{R}_k\right) \qquad k = 1,2 \quad \text{(B8b)}$$

$$\frac{\partial {\boldsymbol{\rho}_{1(k)}^{(12)}}^2(Z_{12}^*, Z_{21})}{\partial Z_{j+M/2}^*} = 0 \qquad k = 1,2 \quad \text{(B8c)}$$

$$\frac{\partial {\boldsymbol{\rho}_{1(k)}^{(12)}}^2(Z_{12}^*, Z_{21})}{\partial Z_j} = 0 \qquad k = 1,2 \quad \text{(B8d)}$$

$$\frac{\partial {\boldsymbol{\rho}_{1(k)}^{(12)}}^2(Z_{12}^*, Z_{21})}{\partial Z_{j+M/2}} = \frac{\partial}{\partial Z_{j+M/2}}\sum_{j=1}^{M/2}\left(\frac{Z_j^* + Z_{j+M/2}}{\sqrt{2\gamma}} - \boldsymbol{R}_k\right)^2 = \sqrt{\frac{2}{\gamma}}\left(\frac{Z_j^* + Z_{j+M/2}}{\sqrt{2\gamma}} - \boldsymbol{R}_k\right) \qquad k = 1,2 \quad \text{(B8e)}$$

$$\boldsymbol{\rho}_{2(k)}^{(12)}(Z_{12}^*, Z_{21}) = \frac{Z_{e2}^* + Z_{e1}}{\sqrt{2\gamma}} - \boldsymbol{R}_k \qquad k = 1,2 \quad \text{(B8f)}$$

$$\frac{\partial {\boldsymbol{\rho}_{2(k)}^{(12)}}^2(Z_{12}^*, Z_{21})}{\partial Z_j^*} = \frac{\partial}{\partial Z_j^*}\sum_{j=1}^{M/2}\left(\frac{Z_{j+M/2}^* + Z_j}{\sqrt{2\gamma}} - \boldsymbol{R}_k\right)^2 = 0 \qquad k = 1,2 \quad \text{(B8g)}$$

$$\frac{\partial {\boldsymbol{\rho}_{2(k)}^{(12)}}^2(Z_{12}^*, Z_{21})}{\partial Z_{j+M/2}^*} = \frac{\partial}{\partial Z_{j+M/2}^*}\sum_{j=1}^{M/2}\left(\frac{Z_{j+M/2}^* + Z_j}{\sqrt{2\gamma}} - \boldsymbol{R}_k\right)^2 = \sqrt{\frac{2}{\gamma}}\left(\frac{Z_{j+M/2}^* + Z_j}{\sqrt{2\gamma}} - \boldsymbol{R}_k\right) \qquad k = 1,2 \quad \text{(B8h)}$$

$$\frac{\partial \boldsymbol{\rho}_{2(k)}^{(12)^2}(Z_{12}^*, Z_{21})}{\partial Z_j} = \frac{\partial}{\partial Z_j} \sum_{j=1}^{M/2} \left(\frac{Z_{j+M/2}^* + Z_j}{\sqrt{2\gamma}} - R_k\right)^2 = \sqrt{\frac{2}{\gamma}} \left(\frac{Z_{j+M/2}^* + Z_j}{\sqrt{2\gamma}} - R_k\right) \qquad k = 1,2 \tag{B8i}$$

$$\frac{\partial \boldsymbol{\rho}_{2(k)}^{(12)^2}(Z_{12}^*, Z_{21})}{\partial Z_{j+M/2}} = \frac{\partial}{\partial Z_{j+M/2}} \sum_{j=1}^{M/2} \left(\frac{Z_{j+M/2}^* + Z_j}{\sqrt{2\gamma}} - R_k\right)^2 = 0 \qquad k = 1,2 \tag{B8j}$$

$$\boldsymbol{\rho}_{12}^{(12)}(Z_{12}^*, Z_{21}) = \frac{Z_{e1}^* + Z_{e2}}{\sqrt{2\gamma}} - \frac{Z_{e2}^* + Z_{e1}}{\sqrt{2\gamma}} \tag{B8k}$$

$$\frac{\partial \boldsymbol{\rho}_{12}^{(12)^2}(Z_{12}^*, Z_{21})}{\partial Z_j^*} = \frac{\partial}{\partial Z_j^*} \sum_{j=1}^{M/2} \left(\frac{Z_j^* + Z_{j+M/2}}{\sqrt{2\gamma}} - \frac{Z_{j+M/2}^* + Z_j}{\sqrt{2\gamma}}\right)^2 = \frac{1}{\gamma}\left(Z_j^* + Z_{j+M/2} - Z_{j+M/2}^* - Z_j\right) \tag{B8l}$$
$$= -\frac{\partial \boldsymbol{\rho}_{12}^{(12)^2}(Z_{12}^*, Z_{21})}{\partial Z_{j+M/2}^*}$$

$$\frac{\partial \boldsymbol{\rho}_{12}^{(12)^2}(Z_{12}^*, Z_{21})}{\partial Z_j} = \frac{\partial}{\partial Z_j} \sum_{j=1}^{M/2} \left(\frac{Z_j^* + Z_{j+M/2}}{\sqrt{2\gamma}} - \frac{Z_{j+M/2}^* + Z_j}{\sqrt{2\gamma}}\right)^2 = -\frac{1}{\gamma}\left(Z_j^* + Z_{j+M/2} - Z_{j+M/2}^* - Z_j\right) \tag{B8m}$$
$$= -\frac{\partial \boldsymbol{\rho}_{12}^{(12)^2}(Z_{12}^*, Z_{21})}{\partial Z_{j+M/2}}.$$

Therefor $Z^*$ derivatives of $\widetilde{H}(Z_{12}^*, Z_{21})$ can be derived as follows

$$\frac{\partial \widetilde{H}(Z_{12}^*, Z_{21})}{\partial Z_{j,j+M/2}^*} = -\frac{\gamma}{2}\left(Z_{j,j+M/2}^* - Z_{j+M/2,j}\right) \tag{B9}$$
$$- \frac{\partial \boldsymbol{\rho}_{1(1)}^{(12)^2}(Z_{12}^*, Z_{21})}{\partial Z_{j,j+M/2}^*} \left(\frac{-erf\left(\sqrt{\gamma \boldsymbol{\rho}_{1(1)}^{(12)^2}(Z_{12}^*, Z_{21})}\right)}{2\sqrt{\boldsymbol{\rho}_{1(1)}^{(12)}(Z_{12}^*, Z_{21})}^3} + \sqrt{\frac{\gamma}{\pi}} \frac{exp\left(-\gamma \sqrt{\boldsymbol{\rho}_{1(1)}^{(12)}(Z_{12}^*, Z_{21})}^2\right)}{\sqrt{\boldsymbol{\rho}_{1(1)}^{(12)}(Z_{12}^*, Z_{21})}^2}\right)$$
$$- \frac{\partial \boldsymbol{\rho}_{1(2)}^{(12)^2}(Z_{12}^*, Z_{21})}{\partial Z_{j,j+M/2}^*} \left(\frac{-erf\left(\sqrt{\gamma \boldsymbol{\rho}_{1(2)}^{(12)^2}(Z_{12}^*, Z_{21})}\right)}{2\sqrt{\boldsymbol{\rho}_{1(2)}^{(12)}(Z_{12}^*, Z_{21})}^3} + \sqrt{\frac{\gamma}{\pi}} \frac{exp\left(-\gamma \sqrt{\boldsymbol{\rho}_{1(2)}^{(12)}(Z_{12}^*, Z_{21})}^2\right)}{\sqrt{\boldsymbol{\rho}_{1(2)}^{(12)}(Z_{12}^*, Z_{21})}^2}\right)$$
$$- \frac{\partial \boldsymbol{\rho}_{2(1)}^{(12)^2}(Z_{12}^*, Z_{21})}{\partial Z_{j,j+M/2}^*} \left(\frac{-erf\left(\sqrt{\gamma \boldsymbol{\rho}_{2(1)}^{(12)^2}(Z_{12}^*, Z_{21})}\right)}{2\sqrt{\boldsymbol{\rho}_{2(1)}^{(12)}(Z_{12}^*, Z_{21})}^3} + \sqrt{\frac{\gamma}{\pi}} \frac{exp\left(-\gamma \sqrt{\boldsymbol{\rho}_{2(1)}^{(12)}(Z_{12}^*, Z_{21})}^2\right)}{\sqrt{\boldsymbol{\rho}_{2(1)}^{(12)}(Z_{12}^*, Z_{12})}^2}\right)$$
$$- \frac{\partial \boldsymbol{\rho}_{2(2)}^{(12)^2}(Z_{12}^*, Z_{21})}{\partial Z_{j,j+M/2}^*} \left(\frac{-erf\left(\sqrt{\gamma \boldsymbol{\rho}_{2(2)}^{(12)^2}(Z_{12}^*, Z_{21})}\right)}{2\sqrt{\boldsymbol{\rho}_{2(2)}^{(12)}(Z_{12}^*, Z_{12})}^3} + \sqrt{\frac{\gamma}{\pi}} \frac{exp\left(-\gamma \sqrt{\boldsymbol{\rho}_{2(2)}^{(12)}(Z_{12}^*, Z_{21})}^2\right)}{\sqrt{\boldsymbol{\rho}_{2(2)}^{(12)}(Z_{12}^*, Z_{21})}^2}\right)$$
$$+ \frac{\partial \boldsymbol{\rho}_{12}^{(12)^2}(Z_{12}^*, Z_{21})}{\partial Z_{j,j+M/2}^*} \left(\frac{-erf\left(\sqrt{\frac{\gamma}{2} \boldsymbol{\rho}_{12}^{(12)^2}(Z_{12}^*, Z_{21})}\right)}{2\sqrt{\boldsymbol{\rho}_{12}^{(12)}(Z_{12}^*, Z_{21})}^3} + \sqrt{\frac{\gamma}{2\pi}} \frac{exp\left(-\frac{\gamma}{2} \sqrt{\boldsymbol{\rho}_{12}^{(12)}(Z_{12}^*, Z_{21})}^2\right)}{\sqrt{\boldsymbol{\rho}_{12}^{(12)}(Z_{12}^*, Z_{21})}^2}\right)$$

and similarly, for $Z$ derivatives of $\widetilde{H}(Z_{12}^*, Z_{21})$ we would have

$$\frac{\partial \widetilde{H}(Z_{12}^*, Z_{21})}{\partial Z_{j,j+M/2}} = -\gamma(Z_{j,j+M/2} - Z_{j+M/2,j}^*) \tag{B10}$$

$$-\frac{\partial \rho_{1(1)}^{(12)^2}(Z_{12}^*, Z_{21})}{\partial Z_{j,j+M/2}} \left( \frac{-erf\left(\sqrt{\gamma \rho_{1(1)}^{(12)^2}(Z_{12}^*, Z_{21})}\right)}{2\sqrt{\rho_{1(1)}^{(12)}(Z_{12}^*, Z_{21})}^3} + \sqrt{\frac{\gamma}{\pi}} \frac{exp\left(-\gamma \sqrt{\rho_{1(1)}^{(12)}(Z_{12}^*, Z_{21})}^2\right)}{\sqrt{\rho_{1(1)}^{(12)}(Z_{12}^*, Z_{21})}^2} \right)$$

$$-\frac{\partial \rho_{1(2)}^{(12)^2}(Z_{12}^*, Z_{21})}{\partial Z_{j,j+M/2}} \left( \frac{-erf\left(\sqrt{\gamma \rho_{1(2)}^{(12)^2}(Z_{12}^*, Z_{21})}\right)}{2\sqrt{\rho_{1(2)}^{(12)}(Z_{12}^*, Z_{21})}^3} + \sqrt{\frac{\gamma}{\pi}} \frac{exp\left(-\gamma \sqrt{\rho_{1(2)}^{(12)}(Z_{12}^*, Z_{21})}^2\right)}{\sqrt{\rho_{1(2)}^{(12)}(Z_{12}^*, Z_{21})}^2} \right)$$

$$-\frac{\partial \rho_{2(1)}^{(12)^2}(Z_{12}^*, Z_{21})}{\partial Z_{j,j+M/2}} \left( \frac{-erf\left(\sqrt{\gamma \rho_{2(1)}^{(12)^2}(Z_{12}^*, Z_{21})}\right)}{2\sqrt{\rho_{2(1)}^{(12)}(Z_{12}^*, Z_{21})}^3} + \sqrt{\frac{\gamma}{\pi}} \frac{exp\left(-\gamma \sqrt{\rho_{2(1)}^{(12)}(Z_{12}^*, Z_{21})}^2\right)}{\sqrt{\rho_{2(1)}^{(12)}(Z_{12}^*, Z_{12})}^2} \right)$$

$$-\frac{\partial \rho_{2(2)}^{(12)^2}(Z_{12}^*, Z_{21})}{\partial Z_{j,j+M/2}} \left( \frac{-erf\left(\sqrt{\gamma \rho_{2(2)}^{(12)^2}(Z_{12}^*, Z_{21})}\right)}{2\sqrt{\rho_{2(2)}^{(12)}(Z_{12}^*, Z_{12})}^3} + \sqrt{\frac{\gamma}{\pi}} \frac{exp\left(-\gamma \sqrt{\rho_{2(2)}^{(12)}(Z_{12}^*, Z_{21})}^2\right)}{\sqrt{\rho_{2(2)}^{(12)}(Z_{12}^*, Z_{21})}^2} \right)$$

$$+\frac{\partial \rho_{12}^{(12)^2}(Z_{12}^*, Z_{21})}{\partial Z_{j,j+M/2}^*} \left( \frac{-erf\left(\sqrt{\frac{\gamma}{2} \rho_{12}^{(12)^2}(Z_{12}^*, Z_{21})}\right)}{2\sqrt{\rho_{12}^{(12)}(Z_{12}^*, Z_{21})}^3} + \sqrt{\frac{\gamma}{2\pi}} \frac{exp\left(-\frac{\gamma}{2} \sqrt{\rho_{12}^{(12)}(Z_{12}^*, Z_{21})}^2\right)}{\sqrt{\rho_{12}^{(12)}(Z_{12}^*, Z_{21})}^2} \right).$$

## APPENDIX C: THE DYNAMIC EQUATION OF THE WAVE FUNCTION IN THE FCCS-I METHOD

One can verify that, using the identity operator on the base of SCS, TDSE will be led to

$$\frac{d\langle Z_{sj}|\psi\rangle}{dt} = \langle \dot{Z}_{sj}|\psi\rangle - \frac{i}{\hbar} \sum_{k,l=1}^{N} \langle Z_{sj}|\widehat{H}|Z_{sk}\rangle (\Omega_s^{-1})_{kl} \langle Z_{sl}|\psi\rangle. \tag{C1}$$

Using the identity operator again, we have

$$\frac{d\langle Z_{sj}|\psi\rangle}{dt} = \sum_{k,l=1}^{N} \langle \dot{Z}_{sj}|Z_{sk}\rangle (\Omega_s^{-1})_{kl} \langle Z_{sl}|\psi\rangle - \frac{i}{\hbar} \sum_{k,l=1}^{N} \langle Z_{sj}|\widehat{H}|Z_{sk}\rangle (\Omega_s^{-1})_{kl} \langle Z_{sl}|\psi\rangle. \tag{C2}$$

By substituting $\langle Z_{sj}|\psi\rangle$ by $C_s exp\left(\frac{iS_s}{\hbar}\right)$ and taking time derivative, we will reach to

$$\frac{d\langle Z_{sj}|\psi\rangle}{dt} = \frac{d}{dt}\left(C_{sj} exp\left(\frac{iS_{sj}}{\hbar}\right)\right) = exp\left(\frac{iS_{sj}}{\hbar}\right)\frac{dC_{sj}}{dt} + \frac{i}{\hbar} C_{sj} \frac{dS_{sj}}{dt} exp\left(-\frac{iS_{sj}}{\hbar}\right) \tag{C3}$$

$$\frac{dC_{sj}}{dt} = -\frac{i}{\hbar} C_{sj} \frac{dS_{sj}}{dt} + \frac{d\langle Z_{sj}|\psi\rangle}{dt} exp\left(-\frac{iS_{sj}}{\hbar}\right). \tag{C4}$$

Substitute the Eq. (C2) in the Eq. (C4) then we have

$$\frac{dC_{sj}}{dt} = -\frac{i}{\hbar} C_{sj} \frac{dS_{sj}}{dt} + \sum_{k,l=1}^{N} \left[ \langle \dot{Z}_{sj}|Z_{sk}\rangle (\Omega_s^{-1})_{kl} - \frac{i}{\hbar} \langle Z_{sj}|\widehat{H}|Z_{sk}\rangle (\Omega_s^{-1})_{kl} \right] C_{sl} exp\left(\frac{i(S_{sl} - S_{sj})}{\hbar}\right). \tag{C5}$$

In this equation, we need to compute $\langle \dot{Z}_{sj}|Z_{sk}\rangle$. Using Eqs. (36) - (37) we can write

$$\langle \dot{Z}_{sj}|Z_{sk}\rangle = (\langle Z_{12j}|Z_{sk}\rangle + \langle Z_{12j}|Z_{sk}\rangle) \frac{\partial}{\partial t}\left(2(1+a_j)\right)^{-1/2} + \left(\left\langle Z_{12j}\left|\frac{\overleftarrow{\partial}}{\partial t}\right|Z_{sk}\right\rangle + \left\langle Z_{21j}\left|\frac{\overleftarrow{\partial}}{\partial t}\right|Z_{sk}\right\rangle\right)\left(2(1+a_j)\right)^{-1/2} \tag{C6}$$

$$\langle \dot{Z}_{sj}|Z_{sk}\rangle = -\frac{1}{2(1+a_j)} \left( \frac{\langle Z_{12j}|Z_{sk}\rangle + \langle Z_{21j}|Z_{sk}\rangle}{\left(2(1+a_j)\right)^{1/2}} \right) \left( \sum_{i=1}^{M/2} \left( \frac{\partial a_j}{\partial Z_{ji}^*} \dot{Z}_{ji}^* + \frac{\partial a_j}{\partial Z_{ji+M/2}^*} \dot{Z}_{ji+M/2}^* + \frac{\partial a_j}{\partial Z_{ji}} \dot{Z}_{ji} + \frac{\partial a_j}{\partial Z_{ji+M/2}} \dot{Z}_{ji+M/2} \right) \right) \tag{C7}$$

$$+ \left(\left\langle Z_{12j}\left|\frac{\overleftarrow{\partial}}{\partial t}\right|Z_{sk}\right\rangle + \left\langle Z_{21j}\left|\frac{\overleftarrow{\partial}}{\partial t}\right|Z_{sk}\right\rangle\right)\left(2(1+a_j)\right)^{-1/2}$$

$$\langle \dot{Z}_{sj}|Z_{sk}\rangle = -\frac{\langle Z_{sj}|Z_{sk}\rangle}{2(1+a_j)}\left(\sum_{i=1}^{M/2}\left(\frac{\partial a_j}{\partial Z_{ji}^*}\dot{Z}_{ji}^* + \frac{\partial a_j}{\partial Z_{ji+M/2}^*}\dot{Z}_{ji+M/2}^* + \frac{\partial a_j}{\partial Z_{ji}}\dot{Z}_{ji} + \frac{\partial a_j}{\partial Z_{ji+M/2}}\dot{Z}_{ji+M/2}\right)\right) \quad \text{(C8)}$$

$$+\frac{1}{\sqrt{2(1+a_k)2(1+a_j)}}(\langle \dot{Z}_{e1j}|\langle Z_{e2j}||Z_{e1k}\rangle|Z_{e2k}\rangle + \langle \dot{Z}_{e1j}|\langle Z_{e2j}||Z_{e2k}\rangle|Z_{e1k}\rangle$$
$$+ \langle Z_{e1j}|\langle \dot{Z}_{e2j}||Z_{e1k}\rangle|Z_{e2k}\rangle + \langle Z_{e1j}|\langle \dot{Z}_{e2j}||Z_{e2k}\rangle|Z_{e1k}\rangle + \langle \dot{Z}_{e2j}|\langle Z_{e1j}||Z_{e1k}\rangle|Z_{e2k}\rangle$$
$$+ \langle \dot{Z}_{e2j}|\langle Z_{e1j}||Z_{e2k}\rangle|Z_{e1k}\rangle + \langle Z_{e2j}|\langle \dot{Z}_{e1j}||Z_{e1k}\rangle|Z_{e2k}\rangle + \langle Z_{e2j}|\langle \dot{Z}_{e1j}||Z_{e2k}\rangle|Z_{e1k}\rangle).$$

Using Eqs. (A4a) – (A4d) in (C8) would be led to

$$\langle \dot{Z}_{sj}|Z_{sk}\rangle = -\frac{a_j}{2(1+a_j)}\left(\sum_{i=1}^{M/2}\left((Z_{ji+M/2}-Z_{ji})\dot{Z}_{ji}^* + (Z_{ji}-Z_{ji+M/2})\dot{Z}_{ji+M/2}^* + (Z_{ji+M/2}^*-Z_{ji}^*)\dot{Z}_{ji}\right.\right. \quad \text{(C9)}$$

$$\left.\left.+ (Z_{ji}^*-Z_{ji+M/2}^*)\dot{Z}_{ji+M/2}\right)\right)\langle Z_{sj}|Z_{sk}\rangle$$

$$+\frac{1}{\sqrt{2(1+a_k)2(1+a_j)}}(\langle \dot{Z}_{e1j}|Z_{e1k}\rangle\langle Z_{e2j}|Z_{e2k}\rangle + \langle \dot{Z}_{e1j}|Z_{e2k}\rangle\langle Z_{e2j}|Z_{e1k}\rangle + \langle Z_{e1j}|Z_{e1k}\rangle\langle \dot{Z}_{e2j}|Z_{e2k}\rangle$$
$$+ \langle Z_{e1j}|Z_{e2k}\rangle\langle \dot{Z}_{e2j}|Z_{e1k}\rangle + \langle \dot{Z}_{e2j}|Z_{e1k}\rangle\langle Z_{e1j}|Z_{e2k}\rangle + \langle \dot{Z}_{e2j}|Z_{e2k}\rangle\langle Z_{e1j}|Z_{e1k}\rangle + \langle Z_{e2j}|Z_{e1k}\rangle\langle \dot{Z}_{e1j}|Z_{e2k}\rangle$$
$$+ \langle Z_{e2j}|Z_{e2k}\rangle\langle \dot{Z}_{e1j}|Z_{e1k}\rangle)$$

consider

$$\langle \dot{Z}_{e1j}|Z_{e1k}\rangle\langle Z_{e2j}|Z_{e2k}\rangle = \langle Z_{e2j}|Z_{e2k}\rangle\langle \dot{Z}_{e1j}|Z_{e1k}\rangle \ , \ \langle \dot{Z}_{e1j}|Z_{e2k}\rangle\langle Z_{e2j}|Z_{e1k}\rangle = \langle Z_{e2j}|Z_{e1k}\rangle\langle \dot{Z}_{e1j}|Z_{e2k}\rangle \quad \text{(C10)}$$
$$\langle Z_{e1j}|Z_{e1k}\rangle\langle \dot{Z}_{e2j}|Z_{e2k}\rangle = \langle \dot{Z}_{e2j}|Z_{e2k}\rangle\langle Z_{e1j}|Z_{e1k}\rangle \ , \ \langle Z_{e1j}|Z_{e2k}\rangle\langle \dot{Z}_{e2j}|Z_{e1k}\rangle = \langle \dot{Z}_{e2j}|Z_{e1k}\rangle\langle Z_{e1j}|Z_{e2k}\rangle$$

and use Eqs. (A2e) – (A2h) we can write

$$\langle \dot{Z}_{sj}|Z_{sk}\rangle = -\frac{a_j}{2(1+a_j)}\left(\sum_{i=1}^{M/2}\left((Z_{ji+M/2}-Z_{ji})\dot{Z}_{ji}^* + (Z_{ji}-Z_{ji+M/2})\dot{Z}_{ji+M/2}^* + (Z_{ji+M/2}^*-Z_{ji}^*)\dot{Z}_{ji}\right.\right. \quad \text{(C11)}$$

$$\left.\left.+ (Z_{ji}^*-Z_{ji+M/2}^*)\dot{Z}_{ji+M/2}\right)\right)\langle Z_{sj}|Z_{sk}\rangle$$

$$-\frac{1}{2}\left(\sum_{i=1}^{n_{df}/2}(Z_{ji}\dot{Z}_{ji}^* + Z_{ji}^*\dot{Z}_{ji} + Z_{ji+M/2}\dot{Z}_{ji+M/2}^* + Z_{ji+M/2}^*\dot{Z}_{ji+M/2})\right)\left(2\frac{\langle Z_{12j}|Z_{12k}\rangle + \langle Z_{12j}|Z_{21k}\rangle}{\sqrt{2(1+a_k)2(1+a_j)}}\right)$$

$$+\frac{2}{\sqrt{2(1+a_k)2(1+a_j)}}\left(\langle Z_{12j}|Z_{12k}\rangle \sum_{i=1}^{M/2}(Z_{ki}\dot{Z}_{ji}^* + Z_{ki+M/2}\dot{Z}_{ji+M/2}^*)\right.$$

$$\left.+ \langle Z_{21j}|Z_{12k}\rangle \sum_{i=1}^{M/2}(Z_{ki+M/2}\dot{Z}_{ji}^* + Z_{ki}\dot{Z}_{ji+M/2}^*)\right).$$

The matrix elements of the overlap $\langle Z_{sj}|Z_{sk}\rangle$ can be computed as follows

$$\langle Z_{sj}|Z_{sk}\rangle = \frac{\langle Z_{12j}|Z_{12k}\rangle + \langle Z_{12j}|Z_{21k}\rangle + \langle Z_{21j}|Z_{12k}\rangle + \langle Z_{21j}|Z_{21k}\rangle}{\sqrt{2(1+a_k)2(1+a_j)}} = 2\frac{\langle Z_{12j}|Z_{12k}\rangle + \langle Z_{12j}|Z_{21k}\rangle}{\sqrt{2(1+a_k)2(1+a_j)}}. \quad \text{(C12)}$$

Using this equation into the Eq. (C11) would be led to

$$\langle \dot{Z}_{sj}|Z_{sk}\rangle = -\frac{\langle Z_{sj}|Z_{sk}\rangle}{2(1+a_j)}\sum_{i=1}^{M/2}\left((Z_{ji}^* + a_j Z_{ji+M/2}^*)\dot{Z}_{ji} + (Z_{ji+M/2}^* + a_j Z_{ji}^*)\dot{Z}_{ji+M/2}\right) \quad \text{(C13)}$$

$$-\frac{\langle Z_{sj}|Z_{sk}\rangle}{2(1+a_j)}\sum_{i=1}^{M/2}\left((Z_{ji} + a_j Z_{ji+M/2})\dot{Z}_{ji}^* + (Z_{ji+M/2} + a_j Z_{ji})\dot{Z}_{ji+M/2}^*\right)$$

$$+\frac{2}{\sqrt{2(1+a_k)2(1+a_j)}}\left(\sum_{i=1}^{M/2}\left[(\langle Z_{12j}|Z_{12k}\rangle Z_{ki} + \langle Z_{21j}|Z_{12k}\rangle Z_{ki+M/2})\dot{Z}_{ji}^*\right.\right.$$

$$\left.\left.+ (\langle Z_{12j}|Z_{12k}\rangle Z_{ki+M/2} + \langle Z_{21j}|Z_{12k}\rangle Z_{ki})\dot{Z}_{ji+M/2}^*\right]\right).$$

Form Eq. (A13), we can show that

$$\sum_{i=1}^{M/2}\left((Z_{ji}^* + a_j Z_{ji+M/2}^*)\dot{Z}_{ji} + (Z_{ji+M/2}^* + a_j Z_{ji}^*)\dot{Z}_{ji+M/2}\right) \quad \text{(C14)}$$

$$= \frac{2(1+a_j)}{i\hbar}\left(\frac{dS_{sj}}{dt} + \langle Z_{sj}|\hat{H}|Z_{sj}\rangle\right) + \sum_{i=1}^{M/2}\left((Z_{ji} + a_j Z_{ji+M/2})\dot{Z}_{ji}^* + (Z_{ji+M/2} + a_j Z_{ji})\dot{Z}_{ji+M/2}^*\right).$$

Substituting the Eq. (C14) into the Eq. (C13) would be led to

$$\langle \dot{Z}_{sj}|Z_{sk}\rangle = \langle Z_{sj}|Z_{sk}\rangle\left(\frac{i}{\hbar}\frac{dS_{sj}}{dt} + \frac{i}{\hbar}\langle Z_{sj}|\hat{H}|Z_{sj}\rangle\right) \quad \text{(C15)}$$

$$-\frac{i}{\hbar}\sum_{i=1}^{M/2}\left(\left(\frac{1}{1+a_j}\langle Z_{sj}|Z_{sk}\rangle Z_{ji} + \frac{a_j}{1+a_j}\langle Z_{sj}|Z_{sk}\rangle Z_{ji+M/2}\right.\right.$$

$$\left.-2\left(\frac{\langle Z_{12j}|Z_{12k}\rangle Z_{ki} + \langle Z_{21j}|Z_{12k}\rangle Z_{ki+M/2}}{\sqrt{2(1+a_k)2(1+a_j)}}\right)\right)\frac{\partial\langle Z_{sj}|\hat{H}|Z_{sj}\rangle}{\partial Z_{ji}}$$

$$+\left(\frac{1}{1+a_j}\langle Z_{sj}|Z_{sk}\rangle Z_{ji+M/2} + \frac{a_j}{1+a_j}\langle Z_{sj}|Z_{sk}\rangle\left(\frac{Z_{ji}^* + Z_{ji}}{2}\right)\right.$$

$$\left.\left.-2\left(\frac{\langle Z_{12j}|Z_{12k}\rangle Z_{ki+M/2} + \langle Z_{21j}|Z_{12k}\rangle Z_{ki}}{\sqrt{2(1+a_k)2(1+a_j)}}\right)\right)\frac{\partial\langle Z_{sj}|\hat{H}|Z_{sj}\rangle}{\partial Z_{ji+M/2}}\right).$$

By substituting the Eq. (C15) into the Eq. (C5), one can see that

$$\frac{dC_{sj}}{dt} = -\frac{i}{\hbar}C_{sj}\frac{dS_{sj}}{dt} + \frac{i}{\hbar}\frac{dS_{sj}}{dt}exp\left(-\frac{iS_{sj}}{\hbar}\right)\langle Z_{sj}|\sum_{k,l=1}^{N}|Z_{sk}\rangle(\Omega_s^{-1})_{kl}C_{sl}\,exp\left(\frac{iS_{sl}}{\hbar}\right) \quad \text{(C16)}$$

$$+ \sum_{k,l=1}^{N}\left[\frac{i}{\hbar}\langle Z_{sj}|\hat{H}|Z_{sj}\rangle\langle Z_{sj}|Z_{sk}\rangle\right.$$

$$-\frac{i}{\hbar}\sum_{i=1}^{M/2}\left(\left(\frac{1}{1+a_j}\langle Z_{sj}|Z_{sk}\rangle Z_{ji} + \frac{a_j}{1+a_j}\langle Z_{sj}|Z_{sk}\rangle Z_{ji+M/2}\right.\right.$$

$$\left.\left.-2\left(\frac{\langle Z_{12j}|Z_{12k}\rangle Z_{ki} + \langle Z_{21j}|Z_{12k}\rangle Z_{ki+M/2}}{\sqrt{2(1+a_k)2(1+a_j)}}\right)\right)\frac{\partial\langle Z_{sj}|\hat{H}|Z_{sj}\rangle}{\partial Z_{ji}}\right.$$

$$+\left(\frac{1}{1+a_j}\langle Z_{sj}|Z_{sk}\rangle Z_{ji+M/2} + \frac{a_j}{1+a_j}\langle Z_{sj}|Z_{sk}\rangle\left(\frac{Z_{ji}^* + Z_{ji}}{2}\right)\right.$$

$$\left.-2\left(\frac{\langle Z_{12j}|Z_{12k}\rangle Z_{ki+M/2} + \langle Z_{21j}|Z_{12k}\rangle Z_{ki}}{\sqrt{2(1+a_k)2(1+a_j)}}\right)\right)\frac{\partial\langle Z_{sj}|\hat{H}|Z_{sj}\rangle}{\partial Z_{ji+M/2}}\right)$$

$$\left.-\frac{i}{\hbar}\langle Z_{sj}|\hat{H}|Z_{sk}\rangle\right](\Omega_s^{-1})_{kl}C_{sl}\,exp\left(\frac{i(S_{sl}-S_{sj})}{\hbar}\right).$$

This equation will be easily simplified to

$$\frac{dC_{sj}}{dt} = -\frac{i}{\hbar}C_{sj}\frac{dS_{sj}}{dt} + \frac{i}{\hbar}\frac{dS_{sj}}{dt}exp\left(-\frac{iS_{sj}}{\hbar}\right)\langle Z_{sj}|\sum_{k,l=1}^{N}|Z_{sk}\rangle(\Omega_s^{-1})_{kl}\langle Z_{sl}|\psi\rangle - \frac{i}{\hbar}\sum_{k=1}^{N}\delta_s^2 H\,D_{sk}\,exp\left(\frac{i(S_{sk}-S_{sj})}{\hbar}\right). \quad \text{(C17)}$$

Finally the dynamic equation of the wave function would be derived as follows

$$\frac{dC_{sj}}{dt} = -\frac{i}{\hbar}\sum_{k=1}^{N}\delta_s^2 H\,D_{sk}\,exp\left(\frac{i(S_{sk}-S_{sj})}{\hbar}\right) \quad \text{(C18)}$$

where

$$\delta_s^2 H = \langle Z_{sj}|\hat{H}|Z_{sk}\rangle - \langle Z_{sj}|\hat{H}|Z_{sj}\rangle\langle Z_{sj}|Z_{sk}\rangle \quad \text{(C19)}$$

$$+ \sum_{i=1}^{M/2} \Bigg(\Bigg( \frac{1}{1+a_j}\langle Z_{sj}|Z_{sk}\rangle Z_{ji} + \frac{a_j}{1+a_j}\langle Z_{sj}|Z_{sk}\rangle Z_{ji+M/2}$$

$$- 2\left(\frac{\langle Z_{12j}|Z_{12k}\rangle Z_{ki} + \langle Z_{21j}|Z_{12k}\rangle Z_{ki+M/2}}{\sqrt{2(1+a_k)2(1+a_j)}}\right)\Bigg) \frac{\partial\langle Z_{sj}|\hat{H}|Z_{sj}\rangle}{\partial Z_{ji}}$$

$$+ \Bigg(\frac{1}{1+a_j}\langle Z_{sj}|Z_{sk}\rangle Z_{ji+M/2} + \frac{a_j}{1+a_j}\langle Z_{sj}|Z_{sk}\rangle\left(\frac{Z_{ji}^* + Z_{ji}}{2}\right)$$

$$- 2\left(\frac{\langle Z_{12j}|Z_{12k}\rangle Z_{ki+M/2} + \langle Z_{21j}|Z_{12k}\rangle Z_{ki}}{\sqrt{2(1+a_k)2(1+a_j)}}\right)\Bigg) \frac{\partial\langle Z_{sj}|\hat{H}|Z_{sj}\rangle}{\partial Z_{ji+M/2}}\Bigg).$$